\documentclass[a4paper,fleqn]{cas-dc}

\usepackage{subfigure}
\usepackage[authoryear,longnamesfirst]{natbib}

\begin{document}
\let\WriteBookmarks\relax
\def\floatpagepagefraction{1}
\def\textpagefraction{.001}

\shorttitle{RecField}
\shortauthors{K. Oku}
\title [mode = title]{RecField: Embedding Recommendations into Spatial Movie Organization as Affordances for Serendipitous Encounters}


%

\author[1]{Kenta Oku}[
    orcid=0000-0002-1639-8178
]
\cormark[1]
\ead{okukenta@rins.ryukoku.ac.jp}
\ead[url]{https://okukenta.net/}

\credit{Conceptualization, Methodology, Software, Validation, Formal analysis, Investigation, Data curation, Writing -- original draft, Writing -- review \& editing, Visualization, Project administration}

\affiliation[1]{
  organization={Faculty of Advanced Science and Technology, Ryukoku University},
  addressline={1-5 Yokotani, Seta Oe-cho},
  city={Otsu},
  postcode={520-2194},
  state={Shiga},
  country={Japan}
}



\cortext[1]{Corresponding author}


\begin{abstract}
    Research on serendipity-oriented recommender systems has focused on algorithms that recommend unexpected yet useful items. However, serendipity is increasingly viewed as a user experience shaped by both algorithmic outputs and system features. This study positions recommendations not as the primary goal users pursue but as an environmental affordance that creates opportunities for discovery during other activities. We propose RecField, a spatial movie organization interface where users freely arrange movies into collection zones. For each collection, RecField provides attribute-based recommendation lists labeled according to their recommendation perspectives. A between-subjects experiment ($N$ = 57) examined the effects of recommendations on discoveries made during a foreground task of creating and organizing personal movie profiles. Although recommendations did not significantly affect environment-level perceived serendipity, ratings of encounter-related aspects tended to be higher, with small effect sizes. At the individual-movie level, favorites in the recommendation condition received significantly higher ratings for stimulating the desire to watch and broadening preferences, but not for novelty or unexpectedness. Notably, only 14\% of evaluated favorites in the recommendation condition were added directly from recommendation lists. Differences in ratings showed similar effect sizes even when the analysis was restricted to favorites added through other routes. These findings suggest that embedded recommendations can contribute to serendipitous experiences both directly through item acceptance and indirectly by prompting users' own exploration and re-evaluation. They also highlight the value of evaluating serendipity at both the environment and item levels and the limitations of relying solely on direct recommendation acceptance rates.
\end{abstract}




\date{\today}

\maketitle

\section{Introduction}
\label{sec:intro}

    Recommender systems are widely used in services such as movie, music, book, and news platforms to present users with items matching their preferences from a large pool of candidates.
    Traditional research on recommender systems has focused primarily on accurately predicting users' ratings and behaviors.
    However, the limitations of evaluating these systems solely in terms of predictive accuracy have been widely recognized \citep{Herlocker2004,McNee2006}, prompting interest in beyond-accuracy perspectives, such as diversity, novelty, and serendipity \citep{Ge2010,Castells2015,Kaminskas2016}.
    Among these perspectives, serendipity has attracted particular attention as a means of counteracting the narrowing of content exposure associated with accuracy-oriented recommendations. It refers to encounters with items that are both unexpected and useful to users \citep{Nguyen2014,ChenL2019WWW}.
    Most studies on serendipity-oriented recommender systems have focused on algorithms for recommending serendipitous items and on metrics for measuring them \citep{Kotkov2016,Ziarani2021}.
    In contrast, \citet{Smets2022IntRS} argued that serendipity in recommender systems should not be narrowly regarded as an algorithmic performance metric, but as a user experience influenced by a variety of mechanisms operating outside the algorithm.
    Information behavior research views serendipity not as mere luck but as arising from the interaction between people and their environment; a phenomenon that can be influenced by environmental design to some extent \citep{Race2016,Bjorneborn2017,McCay-Peet2015IPM}.
    Serendipitous discoveries are often unplanned, occurring as people pursue goals other than the fortuitous revelation \citep{Van-Andel1994,Erdelez2016}.
    Recommendations can be designed not as the primary goal that users actively pursue, but as environmental affordances embedded within another ongoing task to offer opportunities for serendipitous encounters.

    People are known to classify information rigorously and arrange it spatially while its classification remains ambiguous, gradually discovering structure from the arrangement \citep{Malone1983,Marshall1994,Shipman1999}.
    Moreover, collections such as folders and albums are fundamental means by which individuals manage information and curate their interests \citep{Karger2004,Whittaker2011}.
    As a user compiles a collection, its organizing criteria (e.g., director or series) implicitly emerge.
    Presenting related items based on these perspectives can be expected to foster encounters with new items aligned with the user's organizational context.
    However, most studies on interactive recommender systems and serendipity-fostering user interfaces (UIs) investigate tools explicitly designed for recommendation or exploration \citep{HeC2016,Afridi2023}, while research on spatial information organization has primarily addressed arranging information that users already possess.
    The role of an embedded recommendation feature in shaping users' serendipitous experiences during the foreground task of organizing has not been sufficiently clarified.

    This study therefore proposes RecField, an interface that enables users to freely place movies on a canvas called the field and organize them into collection zones.
    When a user forms a collection zone in RecField, the system presents recommendation lists of related movies. These lists are based on the similarity between the collected movies and those defined by attributes such as genre, series, production company, cast, director, and score composer.
    To isolate the effect of the recommendation feature itself, the recommendation logic was deliberately kept simple and a label was attached to each recommendation list indicating its key feature (e.g., ``Director: Hayao Miyazaki''), for users to understand why the movies were presented.

    To verify the effect of the recommendation feature in RecField, a between-subjects experiment was conducted that manipulated its presence.
    Following the experimental design proposed by \citet{Smets2022IntRS}, this study focused its evaluation on movies that participants discovered and added to their favorites while performing a foreground task of creating and organizing their own movie profiles.
    Both the serendipity of the entire system environment, based on the scale proposed by \citet{McCay-Peet2015IPM}, and that of individual movies, based on the questionnaire items of \citet{Kotkov2018} were measured.
    Furthermore, interaction logs were used to identify the path by which each favorite was added and analyze movies selected separately from those added via other routes.
    Specifically, the study addresses the following research questions:
    \begin{description}
        \item[RQ1] Does the presence of the recommendation feature in RecField affect the perceived serendipity of the system environment as a whole?
        \item[RQ2] Does the presence of the recommendation feature affect the perceived serendipity of the movies that users add to their favorites?
        \item[RQ3] Is the serendipity effect of the recommendation feature limited to favorite movies added directly from recommendation results, or does it extend to favorite movies added via other routes?
    \end{description}

    The main contributions of this study are as follows:
    \begin{itemize}
        \item   Recommendations are positioned as environmental affordances that offer opportunities for encounters during organizing activities. Based on this perspective, RecField was designed and implemented as a spatial movie organization interface that presents attribute-labeled recommendation lists for each user-created collection. 
        \item   A between-subjects experiment involving a foreground organizing task was conducted to evaluate the effects of the recommendation on perceived serendipity at two levels: the entire system environment and individual favorite movies.
        \item   The findings show that the effects of the recommendation feature are not limited to movies added directly from recommendation lists. Similar effects were observed in ratings of movies added through other pathways, suggesting that recommendations can contribute to serendipitous experiences through indirect routes.
    \end{itemize}

    The remainder of this paper is organized as follows.
    Section~\ref{sec:related} reviews related work and positions this study.
    Section~\ref{sec:recfield} describes the interface and recommendation logic of RecField.
    Section~\ref{sec:evaluation} sets out the experimental method, and Section~\ref{sec:results} presents the results.
    Section~\ref{sec:discussion} discusses the findings, implications, and limitations of this study, and Section~\ref{sec:conclusion} concludes the paper.

\section{Related work}
\label{sec:related}

    This section reviews work related to this study: serendipity in information behavior (Section~\ref{sec:rw_serendipity}), serendipity-oriented recommender systems (Section~\ref{sec:rw_srs}), serendipity-fostering interfaces and interactive recommender systems (Section~\ref{sec:rw_interface}), and spatial information organization and curation (Section~\ref{sec:rw_organization}). Finally, this study is positioned in terms of these previous studies (Section~\ref{sec:rw_position}).

    \subsection{Serendipity in information behavior}
    \label{sec:rw_serendipity}

        Serendipity has long been discussed as the accidental discovery of something one was not looking for \citep{Van-Andel1994}.
        In library and information science, serendipity and information encountering have been studied as forms of information behavior \citep{Erdelez2016,LiuY2022}.
        These studies treat serendipity not as mere luck but as a phenomenon that arises from interactions between people and their environment that can be cultivated to some extent \citep{Race2016}.
        Based on interviews, \citet{Makri2014} identified strategies that people use to increase their own chances of encountering serendipity.
        \citet{Makri2015} also attempted to directly observe the process by which serendipitous discoveries occur in digital information environments.

        Regarding the design of environments that support seren-dipity, \citet{Bjorneborn2017} proposed a framework linking environmental and personal factors.
        This framework identifies three key affordances that environments should provide: diversifiability (enabling encounters with diverse objects), traversability (enabling free movement within the environment), and sensoriability (making objects easy to perceive).
        \citet{McCay-Peet2015IPM} developed a scale measuring the characteristics of digital environments that foster serendipity (e.g., ``trigger-rich'', ``highlights triggers'', ``enables connections'', and ``leads to the unexpected'') and examined how environmental and personal characteristics affect serendipitous experiences.
        Other work has discussed design guidelines for serendipity-fostering tools \citep{Burkell2012}, interaction patterns \citep{Melo2018}, and practitioners' motivations when designing for serendipity \citep{Smets2023}.
        Building on these insights, this study positions the recommendation feature as an environmental affordance that offers opportunities for encounters during users' exploration and organization activities, not as mere algorithmic output.
        In the evaluation, the scale proposed by \citet{McCay-Peet2015IPM} is also used to assess the environment as a whole.
        
    \subsection{Serendipity-oriented recommender systems}
    \label{sec:rw_srs}

        In recommender systems research, the limitations of evaluation focused solely on predictive accuracy have been pointed out \citep{Herlocker2004,McNee2006}.
        In response, beyond-accuracy objectives such as diversity, novelty, coverage, and serendipity have been discussed \citep{Ge2010,Castells2015,Kaminskas2016}.
        Among these, serendipity has attracted attention as a countermeasure against the content narrowing caused by accuracy-oriented recommendation, commonly known as the filter bubble \citep{Nguyen2014,ChenL2019WWW}.
        Research on serendipity-oriented recommender systems has been summarized in several surveys \citep{Kotkov2016,Ziarani2021}.

        Many early studies focused on algorithms for recommending serendipitous items and on metrics for measuring them.
        For example, methods have been proposed for introducing serendipity into content-based recommendation \citep{Iaquinta2008}, for considering diversity, novelty, and serendipity in music recommendation \citep{ZhangYC2012}, and for enhancing serendipity by fusing two items that a user likes \citep{OkuK2011,OkuK2012}.
        Metrics for evaluating the serendipity of recommendation lists have also been proposed \citep{MurakamiT2008,Ge2010}.

        Meanwhile, other studies treat serendipity as something perceived by users and aim to clarify its components and effects through user evaluations.
        \citet{Kotkov2018} defined serendipity as a combination of relevance, novelty, and unexpectedness.
        They then surveyed real users of the movie recommendation service MovieLens and showed that serendipitous items are associated with the broadening of user preferences.
        Through a large-scale user survey, \citet{ChenL2019WWW} showed that novelty and unexpectedness affect user satisfaction via perceived serendipity.
        \citet{WangN2020} analyzed the effects of item features and users' personal characteristics on perceived serendipity.
        More recently, there have been studies revisiting the definition of serendipity in recommender systems in light of its original meaning of ``fortunate discovery'' \citep{Kotkov2023,Kotkov2024CHIIR}.
        \citet{Smets2022IntRS} argued that serendipity in recommender systems should not be narrowly regarded as an algorithmic performance metric, but rather understood as a user experience influenced by a variety of mechanisms operating outside the algorithm.
        They also classified the mechanisms that foster serendipity and proposed an experimental design for evaluating them.

        As in \citeauthor{Smets2022IntRS}, this study takes the position that serendipity is a user experience.
        The recommendation algorithm is deliberately kept simple and transparent and the effect of embedding the recommendation feature in the interface itself is examined.
        For evaluating individual items, the questionnaire items of \citet{Kotkov2018} are used.

    \subsection{Serendipity-fostering interfaces and interactive recommender systems}
    \label{sec:rw_interface}

        In contrast to conventional recommendation, which unilaterally presents a ranked list, research has sought to support recommendation and exploration through interaction between users and systems.

        \paragraph{Serendipity-fostering interfaces} $\\$

            \citet{Afridi2023} surveyed research on user interfaces that foster serendipity and organized their design patterns.
            The Bohemian Bookshelf by \citet{Thudt2012} provides multiple entry points into a book collection through several coordinated visualizations, supporting serendipitous book discovery.
            \citet{Dias2014HCI} proposed a system that encourages the rediscovery of songs not listened to for a while by visualizing personal music collections.
            Other proposals include a browser that highlights information on the web page being viewed based on the user's own past experiences \citep{Hangal2012}, a faceted browser for Linked Open Data \citep{Khalili2017}, collaborative exploration of book repositories on a tabletop \citep{Radle2012}, and an exploration interface for large-scale video archives \citep{YangY2023}.
            In recommender systems, interfaces that let users slide-adjust recommendations' degrees of serendipity \citep{Afridi2018a} and methods for visualizing serendipitous recommendations \citep{Afridi2018b} have also been investigated.

        \paragraph{Interactive recommender systems} $\\$

            Interactive recommender systems emphasize improving the transparency of recommendations and users' controllability over them \citep{HeC2016,Parra2022}.
            TasteWeights by \citet{Bostandjiev2012} is a hybrid recommender system that enables users to interactively adjust the weights of multiple information sources.
            \citet{Verbert2014} showed that visualizing set relationships among recommendations encourages user exploration.
            Moodplay \citep{Andjelkovic2016} is an interface that provides recommendations while users explore music in a mood space.
            \citet{TsaiCH2018} proposed a visual interface that goes beyond ranked lists to support user-driven exploration and diversification.
            Other studies present recommendations by arranging items on a two-dimensional map or in a three-dimensional space \citep{Gansner2009,Kunkel2017}.
            In the movie domain, MovieExplorer \citep{Taijala2018} enables users to explore movies based on a latent preference space derived from rating data.
            There has also been an attempt to let MovieLens users adjust the characteristics of recommendation algorithms themselves \citep{Harper2015RecSys}.
            The relationship between the degree of user control and users' personal characteristics has also been analyzed \citep{JinY2018RecSys}.
            Being able to understand the reasons for recommendations is an important factor in user trust and acceptance \citep{Tintarev2015RSH}.
            Accordingly, methods that render recommendation results explainable and steerable have also been proposed \citep{Green2009}.

        \paragraph{Presentation formats for recommendations} $\\$

            Designs that group recommendations by category or perspective, rather than presenting them as a single list, have also been investigated.
            \citet{ChenL2010} proposed an organization interface that groups recommended items by trade-off perspectives and demonstrated its effectiveness.
            \citet{Starke2022IntRS} compared single-list, grid, and multi-list interfaces and examined their effects on choice overload \citep{Bollen2010}.
            RecField likewise presents multiple recommendation lists, each with an attribute-based label, for each user-created collection.

        $\\$

        Most of these studies target interfaces whose main purpose is for exploration or to obtain recommendations.
        In contrast, this study embeds a recommendation feature in an environment where users have a different goal (a foreground task), namely organizing movies by themselves.
        It further differs in that it examines how the recommendation feature affects the user experience in that environment.

    \subsection{Spatial information organization and curation}
    \label{sec:rw_organization}

        How people organize information has been extensively studied.
        \citet{Malone1983} observed how people organize documents on office desks and found that they arrange information spatially as rigorously classified ``files'' but also in loosely classified ``piles.''
        He also showed that spatial arrangement serves to remind people of tasks to complete.
        In research on spatial hypertext, VIKI \citep{Marshall1994} was proposed to support users in freely arranging information objects on a two-dimensional plane and in letting structure gradually emerge from the arrangement.
        \citet{Shipman1999} indicated the problems of imposing formal structure on users at an early stage and argued the importance of incrementally formalizing informal representations.
        In personal information management, ``collections'' such as folders and albums are seen as fundamental tools for organizing data \citep{Karger2004}.
        A shift from information consumption to curation has also been noted \citep{Whittaker2011}.
        Furthermore, beyond lookup searches for specific information, studies on exploratory search have emphasized the importance of exploration involving learning and investigation \citep{Marchionini2006}.

        However, these studies on information organization focus on how users organize information they already possess.
        Embedding a recommendation feature into the organization process to foster encounters with new items remains underexplored.

    \subsection{Positioning of this study}
    \label{sec:rw_position}

        Based on the above analysis, the positioning of this study is summarized as follows.

        First, whereas most research on serendipity-oriented recommender systems has focused on improving algorithms, this study, as argued by \citet{Smets2022IntRS}, regards serendipity as a user experience that is also influenced by other mechanisms.
        Based on this, the effect of embedding a recommendation feature into a spatial organization interface is examined.

        Second, whereas most research on interactive recommender systems and serendipity-fostering interfaces has aimed at exploration or obtaining recommendations per se, this study focuses on encounters that occur during the foreground task of organizing movies.

        Third, in the evaluation, both the perceived serendipity of the entire environment \citep{McCay-Peet2015IPM} and the same for individual items \citep{Kotkov2018} are measured.
        Moreover, by analyzing items added directly from recommendation lists separately from those added via other routes, the indirect effects of the recommendation feature are clarified.

\section{RecField}
\label{sec:recfield}

        RecField is a web application that embeds a recommendation feature into an environment where users organize movies spatially.
        Users organize their movies by freely placing them on a canvas called the field and grouping related movies into collection zones.
        Once a collection zone is formed, recommendation lists are presented for each collection based on the attributes of the movies it contains.
        Differently expressed, in RecField, recommendations are not the primary goal that users actively pursue but are positioned as affordances that offer opportunities for encounters during the foreground task of organizing.

\begin{figure*}[t]
	\centering
	\scalebox{1.00}{
		\includegraphics{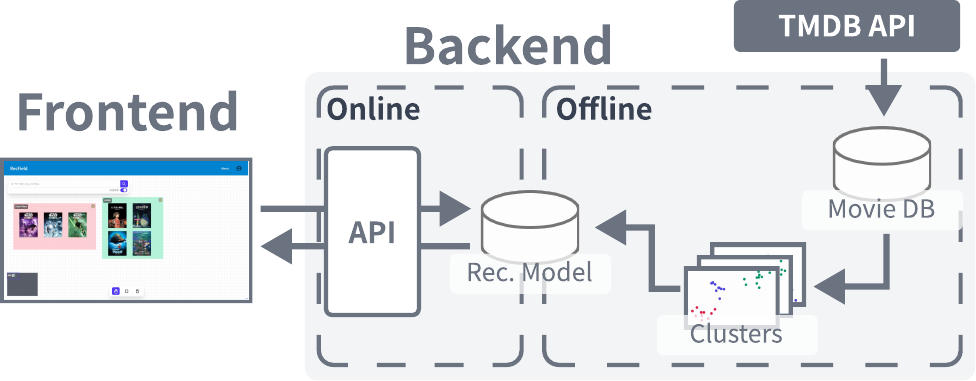}
	}
	\caption{
        Overall architecture of the proposed system, consisting of a frontend, an online recommendation component, and an offline data processing component.
	}
	\label{fig:system_configuration}
\end{figure*}

        Figure~\ref{fig:system_configuration} shows the system configuration of RecField.
        The system consists of a frontend and a backend, with the backend divided into online and offline components.
        The offline component retrieves movie information from the TMDB API and stores it in a movie database.
        Based on this information, movie clusters are generated and provided to the recommendation model.
        The online component handles interactions with the frontend through an API and generates recommendations using the recommendation model.
        The frontend provides an interactive interface through which users can organize movies and explore recommendations.
        
    \subsection{Dataset}
    \label{sec:dataset}

        In this study, the target domain is movies, and we use The Movie Database (TMDB)\footnote{\url{https://www.themoviedb.org/} (accessed October 2026)} as the dataset.
        Movie data were collected using the TMDB API\footnote{\url{https://developer.themoviedb.org/docs/} (accessed October 2026)}.
        We collected data on $11,715$ popular movies available through the TMDB API's Popular Movie Lists as of April 24, 2026.
        We also retrieved the attributes listed in Table~\ref{tab:dataset_stats} as movie metadata.
        Table~\ref{tab:dataset_stats} shows the number of unique values and the number of movie--attribute relations for each attribute.

\begin{table}[t]
	\centering
	\caption{
		Dataset statistics.
        \# unique denotes the number of unique values.
        \# relations denotes the number of movie--attribute relations.
	}
	\scalebox{1.00}{
		\begin{tabular}{l r r}
\hline
Attribute & \# unique & \# relations \\
\hline
Movie & 11,715 & - \\
Genre & 19 & 31,361 \\
Series & 1,356 & 4,464 \\
Company & 10,896 & 35,494 \\
Cast & 46,723 & 112,274 \\
Director & 5,931 & 13,031 \\
Composer & 3,206 & 10,944 \\
\hline

\end{tabular}

	}
	\label{tab:dataset_stats}%
\end{table}%

    \subsection{Interface and interaction design}
    \label{sec:interface}

\begin{figure*}[t]
	\centering
	\scalebox{1.00}{
		\includegraphics{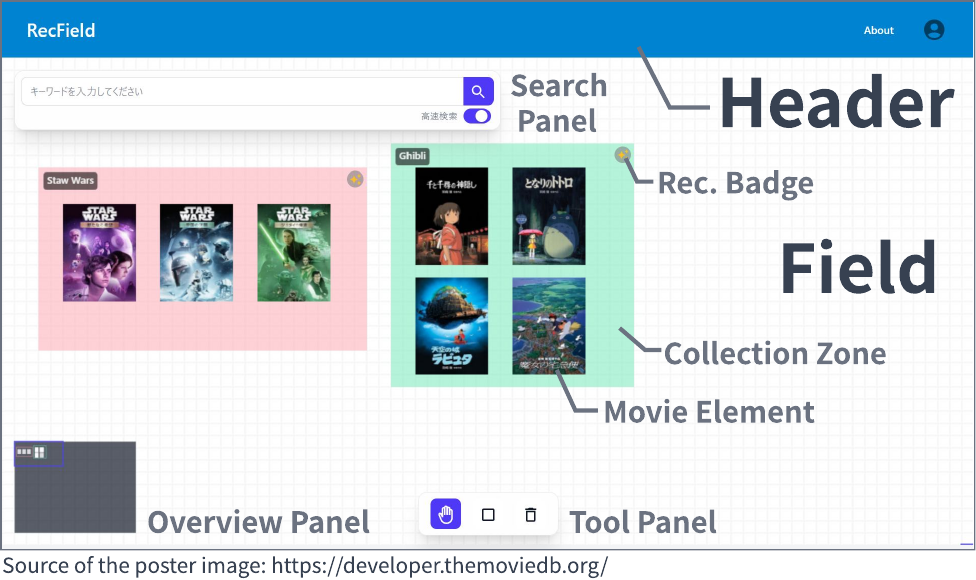}
	}
	\caption{
		Main page of RecField.
        It consists of a header and a field.
        Users can freely organize their favorite movies by placing movies and creating zones on the field.
	}
	\label{fig:page_main}
\end{figure*}

        Figure~\ref{fig:page_main} shows the main page of RecField.
        It comprises a \emph{header} and a canvas called the \emph{field}.
        A \emph{search panel} is located at the top of the field.
        From this search panel, users can search for movies and import them into the field.
        A movie placed on the field is called a \emph{movie element}.
        An area shown as a colored rectangle is called a \emph{collection zone}.
        Users can organize movies by freely arranging these movie elements and collection zones on the field.
        A \emph{tool panel} is located at the bottom of the field.
        The tool panel contains icons for the selection, collection, and deletion tool.
        An \emph{overview panel} at the lower left displays a map of the entire field so that users can grasp the overall layout.

\begin{figure*}[t]
	\centering
	\subfigure[
        Search panel.
        It is displayed on the main page; executing a search displays a search list.
        Clicking the import button of a movie in the search list places that movie on the field.
    ] {
		\centering
		\scalebox{0.55}{
		      \includegraphics{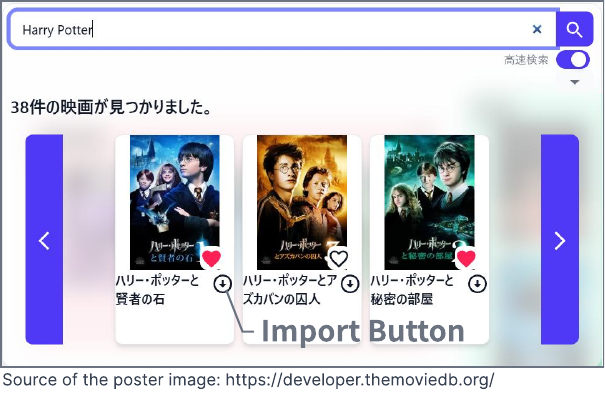}
		}
		\label{fig:panel_search}
	} \subfigure[
        Zone panel.
        It is displayed by clicking a zone.
        Movies related to the movie collection grouped in the zone are presented as recommendation lists.
    ] {
		\centering
		\scalebox{0.55}{
			\includegraphics{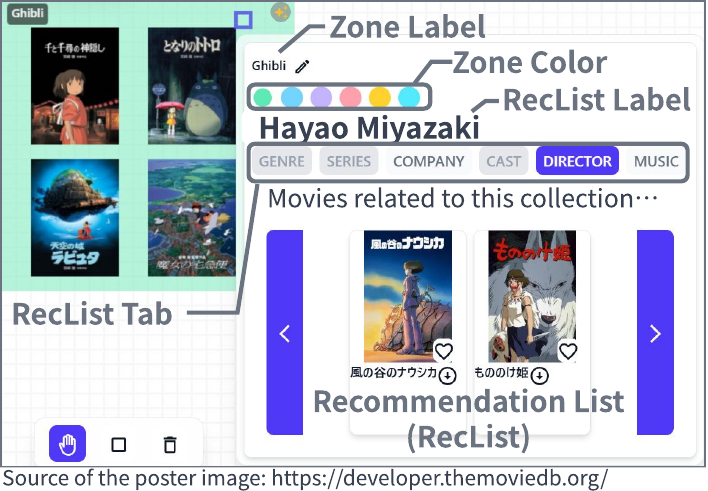}
		}
		\label{fig:panel_zone}
	} \subfigure[
        Movie info panel.
        It is displayed by clicking a movie on the field, or in the search, or a recommendation list.
        Users can view the movie's metadata.
    ] {
		\centering
		\scalebox{0.55}{
			\includegraphics{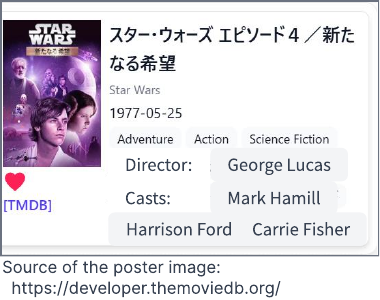}
		}
		\label{fig:panel_movie}
	}
	\caption{
		Panels on the field.
        Each panel is displayed in response to user operations on the field.
	}
	\label{fig:panels}
\end{figure*}

        Figure~\ref{fig:panels} shows the panels displayed on the field.
        Figure~\ref{fig:panel_search} shows the search panel.
        When the user executes a search, the panel expands to display a search results list.
        By clicking the import button attached to a movie card in the search list, the user can place the movie on the field as a movie element.
        Clicking a collection zone on the field displays the zone panel shown in Figure~\ref{fig:panel_zone}.
        The zone panel displays movies related to the collection as a \emph{recommendation list} (RecList), along with a \emph{RecList label} indicating the content of the list.
        In the example in Figure~\ref{fig:panel_zone}, the user has formed a collection of Ghibli movies, and movies directed by Hayao Miyazaki are presented as a recommendation list.
        In this example, recommendation lists are available for the director criterion but also for ``company: Studio Ghibli'' and ``music: Joe Hisaishi.''
        Users can switch between these recommendation perspectives by selecting a RecList tab.
        Users can also edit the zone label and set the zone color as they like.
        Clicking a movie element on the field or a movie card in the search list or a recommendation list displays the movie info panel shown in Figure~\ref{fig:panel_movie}.
        In the movie info panel, users can view the metadata of the movie.
        By clicking the [TMDB] link, users can view the movie's detail page on the TMDB website.
        Clicking the heart button adds the movie to the user's favorites.
        Movies can also be added to favorites from the search list and recommendation lists.

        Table~\ref{tab:user_interactions} lists the user interactions available on the field in RecField.
        Using these interactions, users can freely organize movies by placing movie elements and creating collection zones on the field.

\begin{table*}[t]
	\centering
	\caption{
		User interactions on the field in RecField.
	}
	\scalebox{1.00}{
		\begin{tabular}{p{2cm} p{3cm} p{11cm}}
\hline
Target & Interaction & Description \\
\hline
Search panel & Search for movies & Users can search for movies whose metadata contain a keyword by entering it in the text box. Searches cover not only movie titles but also directors, cast, and other metadata. \\
\hline
Tool panel & Select a tool & Users can select the selection tool, the collection tool, or the deletion tool as the tool for operating on the field. \\
\hline
Movie elements & Import a movie & Clicking the import button of a movie displayed in search results or elsewhere places the movie at an arbitrary position on the field. \\
 & Select a movie & Clicking a movie element selects it. Information on the selected movie is displayed in the movie info panel. \\
 & Move a movie & With the selection tool selected, users can move a movie element by dragging and dropping it. \\
 & Delete a movie & With the deletion tool selected, users can delete a movie element by double-clicking it. \\
 & Add to favorites & Clicking the favorite button adds the movie to the favorites list. \\
\hline
Zones & Create a zone & With the collection tool selected, users can create a collection zone by dragging and dropping on the field. \\
 & Place movie elements in a zone & Users can form a movie collection by placing movie elements in a created zone. \\
 & Select a zone & Clicking a zone selects it. Information on the selected zone is displayed in the zone panel. \\
 & Move or resize a zone & With the selection tool selected, users can move or resize a zone by dragging and dropping it. \\
 & Delete a zone & With the deletion tool selected, users can delete a zone by double-clicking it. \\
\hline
Zone panel & Label a zone & Users can label the target zone by editing its label name. \\
 & Color a zone & Users can change the background color of the zone by selecting a color. \\
 & Switch recommendation lists & Users can switch between recommendation lists by clicking an attribute tab. \\
\hline

\end{tabular}

	}
	\label{tab:user_interactions}%
\end{table*}%

    \subsection{Recommendation logic}
    \label{sec:recommendation}

        When a user places multiple movie elements in a collection zone, the characteristics of that collection are automatically inferred.
        Based on these characteristics, related movies are presented in the form of recommendation lists.
        RecField adopts a simple recommendation logic that emphasizes transparency so that users can clearly understand why each item was recommended in response to their actions.
        For example, when a user forms a collection of movies directed by Steven Spielberg, his movies are presented as a recommendation list; when a user forms a collection of movies from the Harry Potter series, these movies are presented.
        Each recommendation list is given a label indicating its perspective.
        This enables users not only to accept recommendations as they are, but also to reconsider the perspective of their own collection or switch to a different perspective.

        Collection-based recommendation is performed through the following steps:
        \begin{enumerate}[(1)]
            \item   All items are clustered in advance from multiple perspectives (Section~\ref{sec:clustering}).
                    At the same time, each extracted cluster is given a label representing its content.
            \item   When a user forms a collection, all clusters containing every movie in the collection are extracted as candidate clusters. For each candidate cluster, the movies not already included in the collection are used as recommendation candidates (Section~\ref{sec:reclist}).
            \item   For each candidate cluster, the recommendation candidates are presented as a recommendation list for the collection.
        \end{enumerate}

        This section describes the recommendation logic.
        Table~\ref{tab:notation} summarizes the notation used.
        
\begin{table}[t]
	\centering
	\caption{
		Notation.
	}
	\scalebox{0.80}{
		\begin{tabular}{l p{7cm}}
\hline
Notation & Description \\
\hline
$I$ & Set of movies. \\
$i \in I$ & Movie index. \\
$\pi \in \Pi$ & Movie attribute type ($\textrm{genre}$, $\textrm{series}$, \textrm{company}, $\textrm{cast}$, $\textrm{director}$, $\textrm{composer}$). \\
$V^{\pi}$ & Set of values that attribute $\pi$ can take. \\
$V^{\pi}_{i} \subseteq V^{\pi}$ & Set of values of attribute $\pi$ assigned to movie $i \in I$. \\
$\mathcal{C}^{\pi} = \{C_{j}^{\pi}\}_{j=1}^{|\mathcal{C}^{\pi}|}$ & Set of clusters for attribute $\pi$. \\
$I^{\pi}_{j}$ & Set of movies belonging to cluster $C^{\pi}_{j}$. \\
$S_{j} \subseteq V^{\mathrm{genre}}$ & Combination of genres corresponding to genre cluster $C^{\mathrm{genre}}_{j}$ ($1 \le |S_{j}| \le D$). $|S_{j}|$ is called the depth of the cluster. \\
$D$ & Maximum number of genres in a genre combination $S_{j}$. \\
$k$ & Collection index. \\
$I_{k} \subseteq I $ & Set of movies placed in collection $k$. \\
$\mathcal{C}^{\pi, *}_{k}$ & Set of candidate clusters for collection $k$ under attribute $\pi$. \\
$\mathcal{L}^{\pi}_{k,l} \subseteq I$ & Recommendation list corresponding to the $l$-th candidate cluster $C^{\pi}_{j_{l}}$. \\
\hline

\end{tabular}

	}
	\label{tab:notation}%
\end{table}%

        \subsubsection{Construction of attribute-based movie clusters}
        \label{sec:clustering}

\begin{table}[t]
	\centering
	\caption{
		Cluster statistics.
        \# clusters denotes the number of constructed clusters, and \# movies denotes the number of unique movies belonging to the clusters.
        \# movies / cluster denotes the distribution of the number of movies per cluster.
	}
	\scalebox{0.80}{
		\begin{tabular}{c r r r r r r}
\hline
\multirow{2}{*}{Attribute} & \multirow{2}{*}{\# clusters} & \multirow{2}{*}{\# movies} & \multicolumn{4}{c}{\# movies / cluster} \\
 &  &  & Min. & Max. & Avg. & Med. \\
\hline
Genre & 1,208 & 11,591 & 3 & 4849 & 76.47 & 10 \\
Series & 622 & 3,097 & 3 & 80 & 4.98 & 4 \\
Company & 2,464 & 10,103 & 3 & 537 & 10.39 & 5 \\
Cast & 9,351 & 11,063 & 3 & 79 & 7.33 & 5 \\
Director & 1,354 & 6,991 & 3 & 34 & 5.54 & 4 \\
Composer & 840 & 7,444 & 3 & 107 & 9.66 & 5 \\
\hline

\end{tabular}

	}
	\label{tab:clusters_stats}%
\end{table}%

            Movie clusters are constructed for each attribute listed in Table~\ref{tab:dataset_stats}.

            Let $I$ be the set of movies.
            Let $\pi \in \Pi = \{\mathrm{genre}, \mathrm{series},\\ \mathrm{company}, \mathrm{cast}, \mathrm{director}, \mathrm{composer}\}$ denote an attribute type, and let $V^{\pi}$ be the set of values that attribute $\pi$ can take.
            The set of values of attribute $\pi$ assigned to movie $i \in I$ is denoted by $V^{\pi}_{i} \subseteq V^{\pi}$.
            Because a movie can belong to multiple genres, cast members, production companies, and so on, $|V^{\pi}_{i}| \ge 2$ may hold.

            Let $\mathcal{C}^{\pi} = \{C^{\pi}_{j}\}_{j=1}^{|\mathcal{C}^{\pi}|}$ be the set of clusters for attribute $\pi$, and let $I^{\pi}_{j} \subseteq I$ be the set of movies belonging to cluster $C^{\pi}_{j}$.
            Because a movie with $|V^{\pi}_{i}| \ge 2$ belongs to multiple clusters simultaneously, the clusters are not mutually exclusive (i.e., overlapping clustering).

            The construction of cluster sets differs between the genre attribute and the other attributes.
            Each is described below.

            \paragraph{Clusters for attributes other than genre} $\\$

                For $\pi \in \Pi \setminus \{\mathrm{genre}\}$, each attribute value corresponds to one cluster.
                Indexing the attribute values as $V^{\pi} = \{v^{\pi}_{j}\}_{j=1}^{|V^{\pi}|}$, the set of movies belonging to cluster $C^{\pi}_{j}$ is defined as
                \begin{equation}
                    I^{\pi}_{j} = \{i \in I \mid v^{\pi}_{j} \in V^{\pi}_{i}\}.
                \end{equation}

                Hence, $|\mathcal{C}^{\pi}| = |V^{\pi}|$. For example, when $\pi = \mathrm{director}$, $I^{\pi}_{j}$ is the set of movies directed by director $v^{\pi}_{j}$.

                The attribute value $v^{\pi}_{j}$ itself is used as the label of cluster $C^{\pi}_{j}$.
                For example, labels such as ``Star Wars Collection'' and ``Harry Potter Collection'' are used for $\pi = \mathrm{series}$, and labels such as ``Steven Spielberg'' and ``Hayao Miyazaki'' are used for $\pi = \mathrm{director}$.

            \paragraph{Clusters for the genre attribute} $\\$

                There are relatively few genres, and many movies belong to each genre.
                Therefore, mapping each attribute value to one cluster would yield only coarse clusters.
                For $\pi = \mathrm{genre}$, a cluster for each combination of genres is constructed.
                Let $D$ be the maximum number of genres in a combination, and index the subsets of $V^{\mathrm{genre}}$ with at least $1$ and at most $D$ elements as $S_{j}$.
                The set of movies belonging to cluster $C^{\mathrm{genre}}_{j}$ is defined as
                \begin{equation}
                    \begin{aligned}
                        I^{\mathrm{genre}}_{j}
                            &= \{ i \in I \mid S_{j} \subseteq V^{\mathrm{genre}}_{i} \} \\
                            &= \bigcap_{v \in S_{j}} \{ i \in I \mid v \in V^{\mathrm{genre}}_{i} \}.
                    \end{aligned}
                \end{equation}

                That is, $I^{\mathrm{genre}}_{j}$ is the set of movies that belong to all the genres in $S_{j}$.
                These movies may also belong to genres other than those in $S_{j}$.
                We call $|S_{j}|$ the depth of the cluster.
                If $S_{j} \subseteq S_{k}$, then $I^{\mathrm{genre}}_{j} \supseteq I^{\mathrm{genre}}_{k}$.
                The cluster set therefore forms a hierarchy based on inclusion, with clusters becoming finer as depth increases.

                The label of cluster $C^{\pi}_{j}$ is a string concatenating the names of the genres $g$ in $S_{j}$, separated by ``$/$''.
                For example, labels such as ``Action / Comedy / Crime / Family'' and ``Horror / Thriller / Crime'' are used.

                $\\$

                Table~\ref{tab:clusters_stats} shows the cluster statistics.
                Clusters with fewer than three movies were excluded.
                For the genre attribute, we set $D = 6$.

        \subsubsection{Generation of recommendation lists}
        \label{sec:reclist}

            Let $I_{k} \subseteq I$ be the set of movies placed in collection $k$.
            Recommendation lists are generated in the following two steps.

            \paragraph{(1) Extraction of candidate clusters} $\\$

                For each attribute $\pi \in \Pi$, the set of candidate clusters $\mathcal{C}^{\pi, *}_{k}$ for collection $k$ under attribute $\pi$ is defined as the set of clusters that contain all movies in $I_{k}$:
                \begin{equation}
                    \label{eq:candidate_clusters}
                    \mathcal{C}^{\pi, *}_{k} = \{ C^{\pi}_{j} \in \mathcal{C}^{\pi} \mid I_{k} \subseteq I^{\pi}_{j} \}.
                \end{equation}
                That is, the candidate clusters correspond to the attribute values (or, for the genre attribute, the genre combinations) shared by all movies placed in collection $k$.
                Multiple candidate clusters may exist for a single attribute $\pi$.
                We thus denote the number of candidate clusters by $n^{\pi}_{k} = |\mathcal{C}^{\pi, *}_{k}|$ and index them as $\mathcal{C}^{\pi, *}_{k} = \{ C^{\pi}_{j_{1}}, C^{\pi}_{j_{2}}, \dots, C^{\pi}_{j_{n^{\pi}_{k}}} \}$.

            \paragraph{(2) Generation of the recommendation lists} $\\$

                A recommendation list is generated for each candidate cluster.
                The recommendation list $\mathcal{L}^{\pi}_{k,l} \subseteq I$ corresponding to the $l$-th candidate cluster $C^{\pi}_{j_{l}}$ ($l = 1, \dots, n^{\pi}_{k}$) is defined as the set of movies in that cluster excluding the movies $I_{k}$ already placed in collection $k$:
                \begin{equation}
                    \label{eq:reclist}
                    \mathcal{L}^{\pi}_{k,l} = I^{\pi}_{j_{l}} \setminus I_{k}.
                \end{equation}

\section{Evaluation}
\label{sec:evaluation}

    This experiment aimed to examine whether the presence of the recommendation feature in RecField affects users' serendipitous discoveries.
    Specifically, we address the following research questions:
    \begin{description}
	   \item[RQ1] Does the presence of the recommendation feature in RecField affect the perceived serendipity of the entire system environment?
	   \item[RQ2] Does the presence of the recommendation feature affect the perceived serendipity of the movies that users add to their favorites?
	   \item[RQ3] Is the serendipity effect of the recommendation feature limited to favorite movies added directly from recommendation results, or does it extend to favorite movies added via other routes?
    \end{description}

    \subsection{Participants}

        Participants were recruited through the following procedure.
        \begin{enumerate}[(1)]
            \item   On Lancers\footnote{\url{https://www.lancers.jp/}}, a Japanese crowdsourcing service, a pre-screening and recruitment questionnaire was posted. Responses from 500 people were obtained.
                    The pre-screening questionnaire collected information regarding demographic attributes such as age and gender, as well as experience with PCs and recommendation services.
            \item   For the user study, the 500 respondents were asked to complete a questionnaire about their movie-watching frequency and habits. Responses from 400 of them were obtained.
            \item   From the 400 respondents, 210 respondents who watched movies ``at least one day a month'' and had watched at least one movie in the past year were selected.
            \item   From the 210 respondents, 80 respondents were randomly selected.
            \item   The 80 selected respondents were randomly assigned to two groups of 40 each (Groups A and B) and asked to participate in the task described in Section~\ref{sec:procedure}.
        \end{enumerate}

\begin{figure*}[t]
	\centering
	\subfigure[Movie-watching frequency] {
		\centering
		\scalebox{1.00}{
			\includegraphics{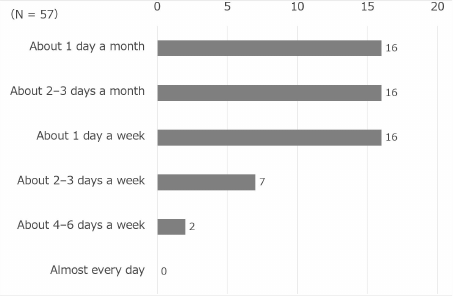}
		}
		\label{fig:participants_movie_frequency}
	} \subfigure[Number of Movies Watched in the Past Year] {
		\centering
		\scalebox{1.00}{
			\includegraphics{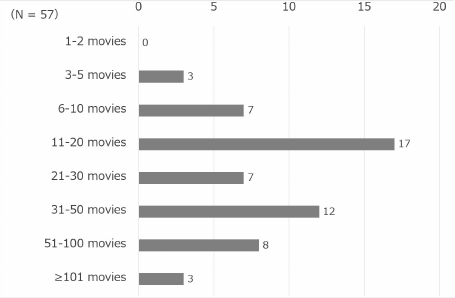}
		}
		\label{fig:participants_movie_amount_yearly}
	}
	\caption{
		Participants' responses regarding their experiences and habits.
	}
	\label{fig:users_experiences}
\end{figure*}

        In total, $N = 57$ participants completed the task: $N_{A} = 29$ in Group A and $N_{B} = 28$ in Group B.
        Of the 57 participants, 43 were men, and 14 were women, aged 26--71 years ($M = 48.4$, $SD = 10.5$).
        Figure~\ref{fig:users_experiences} shows the distributions of participants' responses regarding their experiences and habits.

    \subsection{Procedure}
    \label{sec:procedure}

        This study was reviewed and approved by the Ethics Committee for Research Involving Human Participants at Ryukoku University (approval No. 2025-83). Informed consent was obtained from all participants prior to the experiment.

        Systems for the following two conditions were prepared:
        \begin{description}
	        \item[Condition A] RecField with the recommendation feature
	        \item[Condition B] RecField without the recommendation feature
        \end{description}
        Participants in Group A were assigned to Condition A, and those in Group B to Condition B (between-subjects design).
        Apart from the presence of the recommendation feature, the interface and movie database were identical in both conditions.

        The procedure was based on the experimental design proposed by \citet{Smets2022IntRS}.
        From the perspective that serendipitous discoveries relate to unplanned yet interesting encounters \citep{Bjorneborn2017}, the evaluation was focused on movies that participants discovered while performing a foreground task \citep{Smets2022IntRS}.
        The specific procedure was as follows.

        \begin{enumerate}[(1)]
            \item   Participants were presented with an overview of the study and gave their consent.
            \item   Participants watched a video that explained RecField (approximately 3 minutes).
                    Two versions of the video were prepared, one with and one without the recommendation feature, and each participant was shown the video corresponding to their assigned condition.
            \item   Participants were presented with the following foreground task:
                    \begin{itemize}
                        \item   ``Please use the experimental system to create your own movie profile. Specifically, referring to the explanatory video, choose movies that interest you, such as movies you have watched or would like to watch, and register as many movies as possible while creating collection zones and placing movies in them, organizing them freely.''
                    \end{itemize}
                    Participants were also instructed to add any movies that caught their interest to their favorites while working on the foreground task.
            \item   Participants performed the above task using RecField under their assigned condition.
                    Participants were advised to spend approximately 10 minutes on the task although no time limit was imposed.
            \item   Participants freely selected up to five movies from among their favorites and answered the questions shown in Table~\ref{tab:serendipitous_items_questionnaire} for each movie.
            \item   Finally, participants answered the questions shown in Table~\ref{tab:serendipitous_digital_environment_questionnaire} to evaluate the system environment in its entirety.
        \end{enumerate}

        Item F1 in Table~\ref{tab:serendipitous_items_questionnaire} asks about the participant's viewing experience with the evaluated movie.
        Items F2--F9 were adapted from the questionnaire items of \citet{Kotkov2018} and measure the components of serendipity for individual movies (novelty, unexpectedness, satisfaction, and preference broadening).
        Items E2--E8 in Table~\ref{tab:serendipitous_digital_environment_questionnaire} are seven representative items selected from the 37 items developed by \citet{McCay-Peet2015IPM} and measure the serendipity-fostering characteristics of the system environment in its entirety.
        Item E1 was added by the authors to account for the unique characteristics of RecField.

        In addition, participants' interactions during the experiment were logged together with the event type (e.g., searching for a movie, placing it on the field, adding it to favorites) and the route through which the interaction was performed (a movie on the field, search results, or a recommendation list).
        These logs were used to analyze the routes by which favorites were added in RQ3.

        Conditions were compared using the Mann--Whitney $U$ test, and the effect size was computed as $r = |Z|/\sqrt{N}$.
        For multiple comparisons within each set of items, $p$-values were adjusted using the Holm method, with a significance level of 5\%.
        Because F1 uses a different scale from the other items, it was excluded from the multiple-comparison adjustment.

\begin{table*}[t]
	\centering
	\caption{
        Questionnaire items for individual movies.
        Items F2--F9 were adapted by the author from the questionnaire items of \citet{Kotkov2018}.
        The wording was also modified so that higher scores indicate more positive evaluations in terms of serendipity.
        For F1, participants chose one of \{1: Never watched, 2: Partially watched, 3: Watched from beginning to end\}.
        For F2--F9, participants responded on a five-point Likert scale \{1: Strongly disagree, 2: Disagree, 3: Neither agree nor disagree, 4: Agree, 5: Strongly agree\}.
	}
	\scalebox{0.95}{
		\begin{tabular}{c l l}
\hline
\multicolumn{2}{c}{Measure} & Statement \\
\hline
F1 & Viewing Experience & Have you ever watched this movie? \\
F2 & Awareness Novelty & The first time I heard of this movie was when the system suggested it to me. \\
F3 & Motivational Novelty & Being presented with this movie by the system made me want to watch it (again). \\
F4 & Expectation Unexpectedness & I did not expect that I would be interested in the movie before discovering it. \\
F5 & Discovery Unexpectedness & This is the type of movie I would not normally discover on my own. \\
F6 & Preference Unexpectedness & This movie is different (e.g., in style, genre, or topic) from the movies I usually watch. \\
F7 & Recommendation Unexpectedness & I was surprised that the system picked this movie to present to me. \\
F8 & Satisfaction & I am glad I discovered this movie. \\
F9 & Preference Broadening & Discovering this movie broadened my preferences. \\
\hline

\end{tabular}

	}
	\label{tab:serendipitous_items_questionnaire}%
\end{table*}%

\begin{table*}[t]
	\centering
	\caption{
		Questionnaire items for the system environment as a whole.
        Items E2--E8 are seven representative items selected from the 37 items developed by \citet{McCay-Peet2015IPM}.
        The wording was modified by the author to fit the context of this experiment.
        All items were answered on a five-point Likert scale \{1: Strongly disagree, 2: Disagree, 3: Neither agree nor disagree, 4: Agree, 5: Strongly agree\}.
	}
	\scalebox{0.85}{
		\begin{tabular}{c l l}
\hline
\multicolumn{2}{c}{Measure} & Statement \\
\hline
E1 & Ease of Organization & It is easy to organize movies in this system. \\
E2 & Ease of Exploration & It is easy to explore movies in this system. \\
E3 & Encouragement of In-Depth Examination & This system makes me want to look at movies more closely and learn more about them. \\
E4 & Diversity & This system contains a diverse range of movies. \\
E5 & Richness of Interesting Content & This system contains many movies that I find interesting. \\
E6 & Understanding of Relationships & It is easy to understand the relationships and connections between movies in this system. \\
E7 & Attention to Items of Interest & Movies that interest me are presented in a way that makes them easy to notice in this system. \\
E8 & Unexpected Encounters & I sometimes come across unexpected movies by chance in this system. \\
\hline

\end{tabular}

	}
	\label{tab:serendipitous_digital_environment_questionnaire}%
\end{table*}%

\section{Results}
\label{sec:results}

    This section compares the results of Groups A and B and answers RQ1--RQ3 in turn.

    \subsection{RQ1}

        RQ1 examined whether the presence of the recommendation feature in RecField affects perceived serendipity of the system as a whole.
        Table~\ref{tab:result_system_srdp} shows the results for the eight items of the perceived serendipity scale \citep{McCay-Peet2015IPM}.
        There were 29 valid responses in Group A and 27 in Group B.

        No significant difference between the groups was found for any item, either before or after adjustment (adjusted $p \geq .952$).
        The median for almost every item was 4, indicating that both groups evaluated the system generally positively.

        In terms of effect sizes, however, the direction of the differences showed a consistent pattern across items.
        Group A showed higher means than Group B for ``Unexpected Encounters'' (E8; 3.59 vs.\ 3.07, $r = 0.209$),
        ``Attention to Items of Interest'' (E7; 3.76 vs.\ 3.44, $r = 0.166$),
        and ``Diversity'' (E4; 3.93 vs.\ 3.52, $r = 0.133$).
        In particular, E8, which asks about chance encounters with unexpected movies, had a median of only 3 in Group B and showed the largest effect size among the eight items.
        In contrast, Group B showed slightly higher values for ``Ease of Organization'' (E1; 3.69 vs.\ 3.85) and ``Ease of Exploration'' (E2; 3.59 vs.\ 3.81).

        In summary, the answer to RQ1 is as follows.
        No statistically significant difference was found in the perceived serendipity of the system environment as a whole between conditions with and without the recommendation feature.
        However, the recommendation feature tended to enhance, with small effect sizes, aspects related to opportunities for encounters, such as ``Unexpected Encounters'' and ``Attention to Items of Interest,''
        whereas it tended to slightly reduce aspects related to user-driven operations, such as ``Ease of Organization'' and ``Ease of Exploration.''

\begin{table*}[t]
	\centering
	\caption{
		Serendipity ratings of the environment as a whole.
        $N = 56$ ($N_{A} = 29, N_{B} = 27$).
	}
	\scalebox{1.00}{
		\begin{tabular}{c | c c c | c c c | c c c c c}
\hline
\multirow{2}{*}{Category} & \multicolumn{3}{c|}{A: with recommendations} & \multicolumn{3}{c|}{B: without recommendations} & \multirow{2}{*}{U} & \multirow{2}{*}{Z} & \multirow{2}{*}{r} & \multicolumn{2}{c}{p value} \\
 & Avg. & SD & Med. & Avg. & SD & Med. &  &  &  & not adjusted & adjusted \\
\hline
E1 & 3.69 & 0.93 & 4 & 3.85 & 0.95 & 4 & 351 & -0.737 & 0.098 & 0.467 & 1.000 \\
E2 & 3.59 & 1.12 & 4 & 3.81 & 1.14 & 4 & 341 & -0.869 & 0.116 & 0.390 & 1.000 \\
E3 & 3.69 & 1.07 & 4 & 3.63 & 0.93 & 4 & 417 & 0.445 & 0.059 & 0.663 & 1.000 \\
E4 & 3.93 & 1.00 & 4 & 3.52 & 1.34 & 4 & 450 & 0.996 & 0.133 & 0.323 & 1.000 \\
E5 & 3.62 & 0.90 & 4 & 3.67 & 1.04 & 4 & 366 & -0.451 & 0.060 & 0.659 & 1.000 \\
E6 & 3.72 & 0.84 & 4 & 3.67 & 1.00 & 4 & 400 & 0.138 & 0.018 & 0.897 & 1.000 \\
E7 & 3.76 & 1.09 & 4 & 3.44 & 1.01 & 4 & 464 & 1.243 & 0.166 & 0.217 & 1.000 \\
E8 & 3.59 & 1.02 & 4 & 3.07 & 1.24 & 3 & 484 & 1.568 & 0.209 & 0.119 & 0.952 \\
\hline

\end{tabular}

	}
	\label{tab:result_system_srdp}%
\end{table*}%

\begin{table*}[t]
	\centering
	\caption{
		Serendipity ratings of individual movies.
        The evaluated movies were selected by each participant (up to five) from the movies they added to favorites.
        $N = 198$ ($N_{A} = 108, N_{B} = 90$).
        Bold indicates items with significant differences ($*: p < 0.05, **: p < 0.01$).
	}
	\scalebox{0.95}{
		\begin{tabular}{c | c c c | c c c | c c c r l r l}
\hline
\multirow{2}{*}{Category} & \multicolumn{3}{c|}{A: With recommendations} & \multicolumn{3}{c|}{B: Without recommendations} & \multirow{2}{*}{U} & \multirow{2}{*}{Z} & \multirow{2}{*}{r} & \multicolumn{4}{c}{p value} \\
 & Avg. & SD & Med. & Avg. & SD & Med. &  &  &  & \multicolumn{2}{c}{not adjusted} & \multicolumn{2}{c}{adjusted} \\
\hline
\# favorites & 8.45 & 7.21 & 7 & 7.75 & 5.24 & 7 & 423 & 0.264 & 0.035 & 0.798 &  & - &  \\
F1 & 2.31 & 0.90 & 3 & 2.63 & 0.74 & 3 & 3986 & -2.688 & 0.191 & \textbf{0.007} & \textbf{**} & - &  \\
F2 & 1.97 & 1.55 & 1 & 1.64 & 1.31 & 1 & 5265 & 1.256 & 0.089 & 0.210 &  & 0.839 &  \\
F3 & 4.02 & 1.34 & 5 & 3.61 & 1.30 & 4 & 5975 & 2.929 & 0.208 & \textbf{0.003} & \textbf{**} & \textbf{0.027} & * \\
F4 & 2.22 & 1.46 & 2 & 1.88 & 1.09 & 2 & 5259 & 1.061 & 0.075 & 0.289 &  & 0.839 &  \\
F5 & 2.07 & 1.29 & 2 & 1.80 & 1.07 & 2 & 5329 & 1.256 & 0.089 & 0.210 &  & 0.839 &  \\
F6 & 1.92 & 1.13 & 2 & 1.88 & 1.06 & 2 & 4882 & 0.059 & 0.004 & 0.954 &  & 0.954 &  \\
F7 & 2.18 & 1.29 & 2 & 1.79 & 1.04 & 1 & 5646 & 2.088 & 0.148 & \textbf{0.037} & \textbf{*} & 0.221 &  \\
F8 & 3.63 & 1.20 & 4 & 3.32 & 1.40 & 4 & 5407 & 1.420 & 0.101 & 0.156 &  & 0.780 &  \\
F9 & 3.16 & 1.35 & 3 & 2.61 & 1.30 & 3 & 5971 & 2.838 & 0.202 & \textbf{0.005} & \textbf{**} & \textbf{0.032} & * \\
\hline

\end{tabular}

	}
	\label{tab:result_favorites_srdp}%
\end{table*}%

    \subsection{RQ2}

        RQ2 examined whether the presence of the recommendation feature affects perceived serendipity of the movies that users added to their favorites in the system.

        The numbers of movies added to favorites are shown in the \# favorites row of Table~\ref{tab:result_favorites_srdp}.
        The mean number of favorites per participant was 8.45 in Group A and 7.75 in Group B, with no significant difference ($p = .798$).
        That is, the recommendation feature did not affect the \emph{quantity} of favorites, and the following comparisons can be regarded as comparing the \emph{quality} of approximately equal numbers of favorites.

        Table~\ref{tab:result_favorites_srdp} shows the results of the serendipity ratings collected for each favorite movie.
        The evaluated favorite movies numbered 108 in Group A (28 participants)\footnote{One of the 29 participants did not select any movies for evaluation.} and 90 in Group B (27 participants).
        First, ``Viewing Experience'' was significantly lower in Group A (F1; 2.31 vs.\ 2.63, $p = .007$, $r = 0.191$).
        This indicates that a relatively higher proportion of unwatched movies were added to favorites in Group A.

        Regarding the components of serendipity,
        Group A gave significantly higher ratings for two items:
        ``Motivational Novelty'' (F3; 4.02 vs.\ 3.61, adjusted $p = .027$, $r = 0.208$) and
        ``Preference Broadening'' (F9; 3.16 vs.\ 2.61, adjusted $p = .032$, $r = 0.202$).
        In addition, ``Recommendation Unexpectedness'' (F7; 2.18 vs.\ 1.79) was significant before adjustment ($p = .037$) but not after Holm adjustment (adjusted $p = .221$).
        No significant differences were found for the other items concerning novelty (F2), unexpectedness (F4--F6), and satisfaction (F8), although Group A's means exceeded Group B's on all items.
        Notably, the means for the unexpectedness items (F4--F7) were low, around 2, in both groups, suggesting that most of the favorites added were movies already known to or within the expectations of the users.
        This can be attributed to the fact that the foreground task in this experiment focused on organizing movies, which led participants to work more often with movies they already knew than with new, unknown ones.

        In summary, the answer to RQ2 is as follows.
        The presence of the recommendation feature led to partial differences in serendipity ratings of favorite movies.
        In Group A, favorite movies were rated significantly higher for stimulating the desire to watch and for broadening preferences.
        In contrast, no significant differences were found in the novelty or unexpectedness of the movies themselves.

    \subsection{RQ3}

\begin{table*}[t]
	\centering
	\caption{
		Number of favorite additions per participant by route.
        $N = 57$ ($N_{A} = 29, N_{B} = 28$).
        $n$ denotes the number of participants who added at least one favorite via the route.
        Because RECLIST -> FAVORITE was available only in Group A, no between-group comparison was performed for it.
	}
	\scalebox{0.85}{
		\begin{tabular}{l | r r r r | r r r r | r r r r r r}
\hline
\multirow{2}{*}{Route} & \multicolumn{4}{c|}{A: With recommendations} & \multicolumn{4}{c|}{B: Without recommendations} & \multirow{2}{*}{U} & \multirow{2}{*}{Z} & \multirow{2}{*}{r} & \multicolumn{2}{c}{pvalue} \\
 & Avg. & SD & Med. & n & Avg. & SD & Med. & n &  &  &  & not adjusted & adjusted \\
\hline
a) FIELD $\rightarrow$ FAVORITE & 5.24 & 9.12 & 4 & 23 & 5.21 & 5.90 & 4 & 23 & 384 & -0.354 & 0.047 & 0.730 & 1.000 \\
b) SEARCH $\rightarrow$ FAVORITE & 3.90 & 7.39 & 0 & 12 & 2.71 & 3.94 & 1 & 14 & 391 & -0.261 & 0.035 & 0.800 & 1.000 \\
c) RECLIST $\rightarrow$ FAVORITE & 1.21 & 2.43 & 0 & 10 & - & - & - & - & - & - & - & - & - \\
\hline

\end{tabular}

	}
	\label{tab:result_favorites_route}%
\end{table*}%

\begin{table*}[t]
	\centering
	\caption{
		Serendipity ratings of favorite movies by route of addition.
        Via RECLIST denotes movies added from recommendation lists in Group A, and Via Other denotes movies added via other routes in Group A.
        $N = 108$ ($N_{1} = 15, N_{2} = 93$).
	}
	\scalebox{1.00}{
		\begin{tabular}{c | c c c | c c c | c c c r l r l}
\hline
\multirow{2}{*}{Category} & \multicolumn{3}{c|}{Via RECLIST} & \multicolumn{3}{c|}{Via Other} & \multirow{2}{*}{U} & \multirow{2}{*}{Z} & \multirow{2}{*}{r} & \multicolumn{4}{c}{p value} \\
 & Avg. & SD & Med. & Avg. & SD & Med. &  &  &  & \multicolumn{2}{c}{not adjusted} & \multicolumn{2}{c}{adjusted} \\
\hline
F1 & 2.13 & 0.99 & 3 & 2.34 & 0.89 & 3 & 621 & -0.792 & 0.076 & 0.428 &  & 0.428 &  \\
F2 & 2.33 & 1.72 & 1 & 1.91 & 1.52 & 1 & 807 & 1.168 & 0.112 & 0.243 &  & 1.000 &  \\
F3 & 4.07 & 1.34 & 4 & 4.01 & 1.35 & 5 & 699 & 0.014 & 0.001 & 0.988 &  & 1.000 &  \\
F4 & 2.00 & 1.20 & 2 & 2.26 & 1.50 & 2 & 663 & -0.327 & 0.031 & 0.744 &  & 1.000 &  \\
F5 & 1.73 & 0.80 & 2 & 2.13 & 1.35 & 2 & 628 & -0.663 & 0.064 & 0.507 &  & 1.000 &  \\
F6 & 1.53 & 0.64 & 1 & 1.98 & 1.18 & 2 & 583 & -1.098 & 0.106 & 0.272 &  & 1.000 &  \\
F7 & 2.47 & 1.19 & 2 & 2.13 & 1.30 & 2 & 827 & 1.208 & 0.116 & 0.227 &  & 1.000 &  \\
F8 & 3.87 & 1.25 & 4 & 3.59 & 1.20 & 4 & 825 & 1.194 & 0.115 & 0.232 &  & 1.000 &  \\
F9 & 3.00 & 1.20 & 3 & 3.18 & 1.38 & 3 & 622 & -0.690 & 0.066 & 0.490 &  & 1.000 &  \\
\hline

\end{tabular}

	}
	\label{tab:result_favorites_srdp_reclist}%
\end{table*}%

\begin{table*}[t]
	\centering
	\caption{
		Serendipity ratings of favorite movies added via routes other than recommendation lists.
        Via Other denotes movies added via routes other than recommendation lists in Group A.
        $N = 183$ ($N_{1} = 93, N_{2} = 90$).
        Bold indicates items with significant differences ($*: p < 0.05, **: p < 0.01$).
	}
	\scalebox{1.00}{
		\begin{tabular}{c | c c c | c c c | c c c r l r l}
\hline
\multirow{2}{*}{Category} & \multicolumn{3}{c|}{A: Via Other} & \multicolumn{3}{c|}{B: Without recommendations} & \multirow{2}{*}{U} & \multirow{2}{*}{Z} & \multirow{2}{*}{r} & \multicolumn{4}{c}{p value} \\
 & Avg. & SD & Med. & Avg. & SD & Med. &  &  &  & \multicolumn{2}{c}{not adjusted} & \multicolumn{2}{c}{adjusted} \\
\hline
F1 & 2.34 & 0.89 & 3 & 2.63 & 0.74 & 3 & 3492 & -2.422 & 0.179 & \textbf{0.015} & * & - & \\
F2 & 1.91 & 1.52 & 1 & 1.64 & 1.31 & 1 & 4438 & 0.895 & 0.066 & 0.371 &  & 1.000 &  \\
F3 & 4.01 & 1.35 & 5 & 3.61 & 1.30 & 4 & 5130 & 2.776 & 0.205 & \textbf{0.006} & ** & \textbf{0.039} & * \\
F4 & 2.26 & 1.50 & 2 & 1.88 & 1.09 & 2 & 4547 & 1.079 & 0.080 & 0.281 &  & 1.000 &  \\
F5 & 2.13 & 1.35 & 2 & 1.80 & 1.07 & 2 & 4633 & 1.345 & 0.099 & 0.175 &  & 0.892 &  \\
F6 & 1.98 & 1.18 & 2 & 1.88 & 1.06 & 2 & 4303 & 0.352 & 0.026 & 0.725 &  & 1.000 &  \\
F7 & 2.13 & 1.30 & 2 & 1.79 & 1.04 & 1 & 4743 & 1.670 & 0.123 & 0.095 &  & 0.569 &  \\
F8 & 3.59 & 1.20 & 4 & 3.32 & 1.40 & 4 & 4570 & 1.117 & 0.083 & 0.264 &  & 1.000 &  \\
F9 & 3.18 & 1.37 & 3 & 2.61 & 1.30 & 3 & 5181 & 2.850 & 0.211 & \textbf{0.004} & ** & \textbf{0.035} & * \\
\hline

\end{tabular}

	}
	\label{tab:result_favorites_srdp_other_vs_b}%
\end{table*}%

        RQ2 showed that serendipity ratings of favorite movies were higher in Group A than in Group B.
        However, Group A's favorites include both movies added directly from recommendation results and movies added via other routes, such as search or movies on the field.
        RQ3 therefore examined whether the serendipity effect of the recommendation feature is limited to movies added directly from recommendation results or extends to movies added via other routes.

        \paragraph{Routes of favorite addition} $\\$

            To determine the extent to which the recommendation feature directly contributed to favorite additions, favorite-addition events were aggregated from the interaction logs by route (Table~\ref{tab:result_favorites_route}).
            In RecField, favorites can be added via three routes:
            \begin{enumerate}[a)]
                \item   from movies placed on the field (FIELD),
                \item   from the search list (SEARCH), and
                \item   from recommendation lists (RECLIST; Group A only).
            \end{enumerate}

            The mean numbers of additions per participant were 5.24 (Group A) and 5.21 (Group B) for a) FIELD, and 3.90 and 2.71 for b) SEARCH, with no significant between-group differences for either ($p = .730$ and $.800$, respectively).
            In Group A, the mean number of additions from c) RECLIST was 1.21 per participant (median 0), and 10 of the 29 participants (34\%) added at least one favorite directly from a recommendation list.

            Among the favorite movies evaluated for serendipity, only 15 of the 108 in Group A (14\%, from 7 participants) were added from c) RECLIST; the remainder were added from a) FIELD (58) and b) SEARCH (35).
            That is, even in Group A, approximately 86\% of favorites were added via routes other than recommendation lists, and direct additions from recommendation lists were limited to a subset of users.

        \paragraph{Serendipity ratings by route of addition} $\\$

            Next, the 108 evaluated favorite movies in Group A were categorized into those added from recommendation results (RECLIST; 15 movies) and those added via other routes (Other; 93 movies) and compared (Table~\ref{tab:result_favorites_srdp_reclist}).

            No significant difference was found for any item (unadjusted $p \geq .227$).
            In terms of means, RECLIST movies were rated slightly higher than Other movies for ``Awareness Novelty'' (F2; 2.33 vs.\ 1.91), ``Recommendation Unexpectedness'' (F7; 2.47 vs.\ 2.13), and ``Satisfaction'' (F8; 3.87 vs.\ 3.59).
            In contrast, they were rated lower for ``Discovery Unexpectedness'' (F5; 1.73 vs.\ 2.13) and ``Preference Unexpectedness'' (F6; 1.53 vs.\ 1.98).
            This suggests that, as expected from RecField's design of presenting recommendations based on the movies in a zone, recommended movies tended to be unknown to users but not outside the types of movies they usually prefer.
            However, all effect sizes were small ($r \leq 0.116$), and the number of RECLIST movies was small; thus, this tendency is not conclusive.

            We then compared the movies in Group A added via routes other than recommendation lists (Other; 93 movies) with the movies in Group B (90 movies) (Table~\ref{tab:result_favorites_srdp_other_vs_b}).
            Even when restricted to movies added via routes other than recommendation lists, Group A gave significantly higher ratings than Group B for
            ``Motivational Novelty'' (F3; 4.01 vs.\ 3.61, adjusted $p = .039$, $r = 0.205$) and
            ``Preference Broadening'' (F9; 3.18 vs.\ 2.61, adjusted $p = .035$, $r = 0.211$).
            ``Viewing experience'' (F1) was also significantly lower in Group A (2.34 vs.\ 2.63, $p = .015$, $r = 0.179$).
            These effect sizes were nearly identical to those obtained when comparing all of Group A with Group B (Table~\ref{tab:result_favorites_srdp}; $r = 0.208$ for F3 and $r = 0.202$ for F9).
            Therefore, the between-group differences observed in RQ2 were not driven by the 15 movies added directly from recommendation lists, but appeared to a similar degree in favorites that users added themselves via other routes.

            In summary, the answer to RQ3 is as follows.
            Favorites added directly from recommendation lists accounted for only 14\% of evaluated favorites in Group A, and their serendipity ratings did not differ significantly from those of favorites added via other routes.
            In contrast, the between-group differences in F3 and F9 were observed to a similar degree even when restricted to favorites added via routes other than recommendations.
            That is, the serendipity effect of the recommendation feature was not limited to movies added directly from recommendation lists but extended to movies added via other routes.
            This result suggests that, rather than directly delivering ``completely unexpected movies,'' the recommendation feature prompted users to search for and re-evaluate related movies, thereby indirectly stimulating their desire to watch and broadening their preferences.

\section{Discussion}
\label{sec:discussion}

    \subsection{Findings}
 	      \paragraph{RQ1:}
            No significant difference in the perceived serendipity of the system environment as a whole was found between conditions with and without the recommendation feature.
            Given that the median was 4 for many items in both groups, RecField's basic interaction of creating zones and spatially placing and organizing movies appears, in itself, to have been accepted to a certain degree as an environment that fosters serendipity, regardless of the recommendation feature.
            Nevertheless, the direction of the effect sizes showed a consistent pattern.
            Group A rated items related to \emph{opportunities for encounters} higher, such as ``Unexpected Encounters'' (E8) and ``Attention to Items of Interest'' (E7),
            whereas Group B rated items related to \emph{user-driven operations} slightly higher, such as ``Ease of Organization'' (E1) and ``Ease of Exploration'' (E2).
            This result suggests that, while the recommendation feature provides users with new opportunities, it may be in a partial trade-off with users' sense of organizing and exploring the field according to their own intentions.

 	      \paragraph{RQ2:}
            In the ratings of favorite movies, Group A gave significantly higher ratings for ``Motivational Novelty'' (F3) and ``Preference Broadening'' (F9), and reported significantly less ``Viewing Experience'' (F1).
            In contrast, no significant differences were found for ``Awareness Novelty'' (F2) or unexpectedness (F4--F7).
            Whereas \citet{Kotkov2018} conceptualize serendipity as a combination of relevance, novelty, and unexpectedness, what the recommendation feature enhanced in this experiment was not ``unexpected discovery'' in the strict sense, but rather aspects obtained as \emph{outcomes} of serendipity, namely stimulating the desire to watch and broadening preferences.
            The low unexpectedness ratings in both groups can be attributed to the fact that the foreground task focused on organizing known movies, so participants mainly added known movies to their favorites.
            Differently expressed, rather than facilitating encounters with movies entirely new to users, the recommendation feature can be interpreted as having led users to re-evaluate known movies, or movies close to their existing interests, in the sense of ``I want to watch this again'' or ``my preferences have broadened.''

 	      \paragraph{RQ3:}
            Favorites added directly from recommendation lists accounted for only 14\% of evaluated favorites in Group A, and their ratings did not differ significantly from those of favorites added through other routes.
            In contrast, even when restricted to favorites added via routes other than recommendation lists, the differences between Groups A and B in F3 and F9 were observed with nearly the same effect sizes as in the comparison using all of Group A.
            This result suggests that the serendipity effect of the recommendation feature arose through the route of directly accepting recommended movies, and through an \emph{indirect route} in which the presentation of recommendations prompted users to recall and search for related movies and to re-evaluate movies on the field.
            In addition, movies added from recommendation lists had slightly higher mean ratings for ``Awareness Novelty'' (F2) and ``Recommendation Unexpectedness'' (F7), but slightly lower ratings for ``Discovery Unexpectedness'' (F5) and ``Preference Unexpectedness'' (F6).
            This suggests that, owing to RecField's design of presenting recommendations based on the movies in a zone, recommended movies were \emph{unknown but not outside users' preferred types}, which may be one reason why direct acceptance of recommendations was limited.

    \subsection{Limitations}

        This study has the following limitations.

        First, among the affordances provided by RecField, we manipulated only the presence of the recommendation feature. The effects of other affordances were not examined.
        Besides recommendations, RecField offers several features that can serve as opportunities for encounters, such as free placement of movies on the field, creation and labeling of collection zones, the search panel, and the overview map.
        Because both groups had access to these features in this experiment, whether the median of 4 for many items in both groups in RQ1 was due to the spatial organization environment itself or to individual features could not be determined.
        Moreover, as indicated by Group B's slightly higher ratings for ease of organization and exploration in RQ1, the recommendation feature may interact with other interface elements; thus, caution is needed when attributing observed effects to the recommendation feature alone.

        Second, for the recommendation algorithm, a simple method that emphasizes transparency was adopted so that users can easily understand the reasons for recommendations. The effects of different recommendation algorithms were not examined.
        RecField identifies attribute-based clusters that contain all movies in the collection and presents the remaining movies in each cluster as a recommendation list, without explicitly considering preference estimation accuracy, diversity, or unexpectedness.
        The absence of significant differences in unexpectedness ratings in RQ2---and the tendency in RQ3 for movies added from recommendation lists to be unknown yet still within users' preferred types---may stem from these algorithmic characteristics.
        Consequently, it remains unclear whether similar effects would emerge under alternative recommendation strategies.

        Third, sample size and the unit of analysis also impose limitations.
        With approximately 28 participants per group, this experiment does not have sufficient statistical power to detect small effects.
        In RQ1 in particular, although the direction of the effect sizes showed a consistent pattern, no statistically significant differences were found.
        Verification with larger samples is needed to determine whether these tendencies reflect the effect of the recommendation feature.
        Furthermore, the item-level analyses in RQ2 and RQ3 treated individual movie ratings as independent observations, although multiple movies were evaluated by the same participant. This within-participant dependency may affect the statistical inference. Therefore, the results of these analyses should be interpreted with caution.

    \subsection{Future work}

        Future work includes the following.

        First, the effects of affordances other than the recommendation feature on serendipity should be examined.
        For example, by manipulating the presence of spatial placement and collection zones and comparing the interface with a list-based organization interface, the contribution of the spatial organization environment itself can be clarified.
        It would also be possible to position each feature regarding diversifiability, traversability, and sensoriability as identified by \citet{Bjorneborn2017}, and to examine their interactions with the recommendation feature.

        Second, the effects of different recommendation methods should be examined.
        In addition to the current similarity-based recommendation method, future work will implement recommendation methods that incorporate diversity and unexpectedness and compare their effects on serendipity ratings.
        In particular, the investigation will examine whether presenting movies that are moderately distant from the zone context, while maintaining  transparency through labels indicating the recommendation perspective, can facilitate preference broadening and unexpected discoveries.

        Third, the relationship between users' organizing behavior, such as creating and labeling zones, and serendipity ratings should be analyzed.
        The interaction logs from this experiment indicate a tendency for Group B to perform more operations such as searching and editing zone labels. Clarifying how the presence of the recommendation feature affects users' organizing behavior and how this relates to serendipitous discoveries is an important direction for future work.

        Fourth, experiments will be conducted with larger numbers of participants and the tendencies observed in this study will be verified through analyses treating participants as random effects.
        Recording the sequence of user behaviors from exposure to recommendations to adding movies to favorites will help clarify the indirect process of searching and re-evaluation triggered by recommendations.
        Long-term usage studies will be conducted to examine how continued use of the recommendation feature affects the broadening of user preferences and actual viewing behavior.

        Finally, the applicability of the proposed approach to domains other than movies, such as books, music, and academic papers, will be examined.

\section{Conclusion}
\label{sec:conclusion}

    This study positions recommendations not as the primary goal users actively pursue, but as an environmental affordance for beneficial encounters while another objective is being sought. To facilitate the process, RecField, a spatial movie organization interface, was proposed and investigated.
    In RecField, when users place movies on the field and organize them into collection zones, recommendation lists labeled with their perspectives are presented based on the similarity between the collection and attribute-based movie clusters.
    Through a between-subjects experiment ($N = 57$) that manipulated the presence of the recommendation feature, the serendipity perceived during the foreground task of organizing movies was evaluated at two levels: the system environment as a whole and individual favorite movies.

    The experiment yielded the following findings.
    First, no significant difference was found in the perceived serendipity of the system environment as a whole between conditions with and without the recommendation feature (RQ1).
    However, tendencies with small effect sizes were observed: i) the recommendation group---Group A--- rated items related to opportunities for encounters higher, such as ``Unexpected Encounters'' and ``Attention to Items of Interest,'' whereas ii) the no-recommendation group---Group B--- rated items related to user-driven operations slightly higher, such as ``Ease of Organization'' and ``Ease of Exploration.''
    Second, for individual movies added to favorites, the recommendation group gave significantly higher ratings for stimulating the desire to watch and for broadening preferences, and a higher proportion of unwatched movies were added (RQ2).
    In contrast, no significant differences were found in novelty or unexpectedness, indicating that the recommendation primarily prompted the re-evaluation of known movies, or movies close to existing interests, rather than unexpected discoveries.
    Third, favorites added directly from recommendation lists accounted for only 14\% of the evaluated favorites in the recommendation group, yet the above between-group differences were observed with similar effect sizes even when restricted to favorites added via routes other than recommendation lists (RQ3).

    These results suggest that recommendations embedded in an organizing interface may contribute to serendipitous experiences by leading users to directly accept recommended items and through an indirect route in which the presentation of recommendations prompts users' own searching and re-evaluation.
    This highlights the limitations of evaluating the value of recommendation features solely by direct metrics such as click-through or acceptance rates. It also demonstrates the usefulness of evaluating serendipity-fostering systems at both the environment and item level.

    Nevertheless, this study was based on whether or not the recommendation feature was present. It did not examine the effects of other affordances, such as spatial placement and collection zones.
    In addition, because a simple, transparency-oriented method was used for recommendation, the effects of different recommendation methods remain unclear.
    Furthermore, a sample size of approximately 28 participants per group is insufficient to detect small effects, and the tendencies observed in RQ1 have not been statistically confirmed.
    Future work will further clarify how recommendations embedded in organizing activities foster serendipitous discoveries by examining the effects of affordances other than recommendations. The proposed approach will be compared with recommendation methods that consider diversity and unexpectedness, and the relationship between users' organizing behavior and serendipity ratings will be analyzed. Additionally, larger-scale and longer-term experiments will be conducted, and the approach applied to domains other than movies.

\section*{Acknowledgement}
    This research was conducted as part of the Ryukoku University Special Research Fellow program for the 2026 academic year.
    We would like to thank Editage (www.editage.jp) for English language editing.


\printcredits
	
\section*{Data availability}
    Data will be made available on reasonable request.

\section*{Declaration of generative AI and AI-assisted technologies in the manuscript preparation process}
    During the preparation of this work, the author used ChatGPT (OpenAI) to assist with English translation and text revision. After using this tool, the author reviewed and edited the content as needed and takes full responsibility for the content of the publication.
    
\balance

\bibliographystyle{cas-model2-names}
\bibliography{references}

@ARTICLE{ChenL2010,
  title     = "Experiments on the preference-based organization interface in
               recommender systems",
  author    = "Chen, Li and Pu, Pearl",
  journal   = "ACM Transactions on Computer-Human Interaction",
  publisher = "Association for Computing Machinery (ACM)",
  volume    =  17,
  number    =  1,
  pages     = "1--33",
  month     =  mar,
  year      =  2010,
  doi       = "10.1145/1721831.1721836",
  language  = "en"
}

@ARTICLE{Marchionini2006,
  title     = "Exploratory search: from finding to understanding",
  author    = "Marchionini, Gary",
  journal   = "Communications of the ACM",
  publisher = "Association for Computing Machinery (ACM)",
  volume    =  49,
  number    =  4,
  pages     = "41--46",
  month     =  apr,
  year      =  2006,
  doi       = "10.1145/1121949.1121979",
  language  = "en"
}

@ARTICLE{Shipman1999,
  title     = "Formality considered harmful: Experiences, emerging themes, and
               directions on the use of formal representations in interactive
               systems",
  author    = "Shipman, III, Frank M and Marshall, Catherine C",
  journal   = "Computer Supported Cooperative Work: CSCW: An International
               Journal",
  publisher = "Springer Science and Business Media LLC",
  volume    =  8,
  number    =  4,
  pages     = "333--352",
  month     =  dec,
  year      =  1999,
  doi       = "10.1023/a:1008716330212",
  language  = "en"
}

@INPROCEEDINGS{Marshall1994,
  title     = "{VIKI}: spatial hypertext supporting emergent structure",
  author    = "Marshall, Catherine C and Shipman, III, Frank M and Coombs, James
               H",
  booktitle = "Proceedings of the 1994 ACM European conference on
               Hypermedia technology",
  publisher = "ACM Press",
  address   = "New York, New York, USA",
  pages     = "13--23",
  year      =  1994,
  doi       = "10.1145/192757.192759"
}

@ARTICLE{Malone1983,
  title     = "How do people organize their desks?: Implications for the design
               of office information systems",
  author    = "Malone, Thomas W",
  journal   = "ACM Transactions on Information Systems",
  publisher = "Association for Computing Machinery (ACM)",
  volume    =  1,
  number    =  1,
  pages     = "99--112",
  month     =  jan,
  year      =  1983,
  doi       = "10.1145/357423.357430",
  language  = "en"
}

@INPROCEEDINGS{YangY2023,
  title     = "Latent Wander: An alternative interface for interactive and
               serendipitous discovery of large {AV} archives",
  author    = "Yang, Yuchen and Zhang, Linyida",
  booktitle = "Proceedings of the 5th Workshop on analySis,
               Understanding and proMotion of heritAge Contents",
  publisher = "ACM",
  address   = "New York, NY, USA",
  pages     = "5--12",
  month     =  "2~" # nov,
  year      =  2023,
  doi       = "10.1145/3607542.3617355"
}

@INPROCEEDINGS{Thudt2012,
  title     = "The bohemian bookshelf: supporting serendipitous book discoveries
               through information visualization",
  author    = "Thudt, Alice and Hinrichs, Uta and Carpendale, Sheelagh",
  booktitle = "Proceedings of the SIGCHI Conference on Human
               Factors in Computing Systems",
  publisher = "ACM",
  address   = "New York, NY, USA",
  pages     = "1461--1470",
  month     =  "5~" # may,
  year      =  2012,
  doi       = "10.1145/2207676.2208607"
}

@INPROCEEDINGS{Kotkov2018,
  title     = "Investigating serendipity in recommender systems based on real
               user feedback",
  author    = "Kotkov, Denis and Konstan, Joseph A and Zhao, Qian and
               Veijalainen, Jari",
  booktitle = "Proceedings of the 33rd Annual ACM Symposium on
               Applied Computing",
  pages     = "1341--1350",
  year      =  2018,
  doi       = "10.1145/3167132.3167276"
}

@INPROCEEDINGS{Ge2010,
  title     = "Beyond accuracy: evaluating recommender systems by coverage and
               serendipity",
  author    = "Ge, Mouzhi and Delgado-Battenfeld, Carla and Jannach, Dietmar",
  booktitle = "Proceedings of the fourth ACM conference on
               Recommender systems",
  pages     = "257--260",
  year      =  2010
}

@ARTICLE{Ziarani2021,
  title    = "Serendipity in recommender systems: A systematic literature review",
  author   = "Ziarani, Reza Jafari and Ravanmehr, Reza",
  journal  = "Journal of Computer Science and Technology",
  volume   =  36,
  number   =  2,
  pages    = "375--396",
  year     =  2021,
  doi      = "10.1007/s11390-020-0135-9"
}

@INPROCEEDINGS{ZhangYC2012,
  title     = "Auralist: Introducing serendipity into music recommendation",
  author    = "Zhang, Yuan Cao and Ó Séaghdha, Diarmuid and Quercia, Daniele and
               Jambor, Tamas",
  booktitle = "Proceedings of the fifth ACM international
               conference on Web search and data mining",
  pages     = "13--22",
  year      =  2012,
  doi       = "10.1145/2124295.2124300"
}

@ARTICLE{Whittaker2011,
  title    = "Personal information management: From information consumption to
              curation",
  author   = "Whittaker, Steve",
  journal  = "Annual Review of Information Science and Technology",
  volume   =  45,
  number   =  1,
  pages    = "1--62",
  year     =  2011
}

@INPROCEEDINGS{WangN2020,
  title     = "The impacts of item features and user characteristics on users'
               perceived serendipity of recommendations",
  author    = "Wang, Ningxia and Chen, Li and Yang, Yonghua",
  booktitle = "Proceedings of the 28th ACM Conference on User
               Modeling, Adaptation and Personalization",
  pages     = "266--274",
  year      =  2020,
  doi       = "10.1145/3340631.3394863"
}

@INPROCEEDINGS{Verbert2014,
  title     = "The effect of different set-based visualizations on user
               exploration of recommendations",
  author    = "Verbert, K and Parra, D and Brusilovsky, P",
  booktitle = "Proceedings of the Joint Workshop on
               Interfaces and Human Decision Making in Recommender Systems",
  pages     = "37--44",
  year      =  2014
}

@INPROCEEDINGS{TsaiCH2018,
  title     = "Beyond the ranked list: User-driven exploration and
               diversification of social recommendation",
  author    = "Tsai, Chun-Hua and Brusilovsky, Peter",
  booktitle = "Proceedings of the 23rd International Conference on
               Intelligent User Interfaces",
  pages     = "239--250",
  year      =  2018,
  doi       = "10.1145/3172944.3172959"
}

@INCOLLECTION{Tintarev2015RSH,
  title     = "Explaining recommendations: Design and evaluation",
  author    = "Tintarev, Nava and Masthoff, Judith",
  editor    = "Ricci, Francesco and Rokach, Lior and Shapira, Bracha",
  booktitle = "Recommender Systems Handbook",
  publisher = "Springer",
  edition   =  2,
  pages     = "353--382",
  year      =  2015,
  doi       = "10.1007/978-1-4899-7637-6\_10"
}

@INPROCEEDINGS{Taijala2018,
  title     = "{MovieExplorer}: Building an interactive exploration tool from
               ratings and latent taste spaces",
  author    = "Taijala, Taavi T and Willemsen, Martijn C and Konstan, Joseph A",
  booktitle = "Proceedings of the 33rd Annual ACM Symposium on
               Applied Computing",
  pages     = "1383--1392",
  year      =  2018,
  doi       = "10.1145/3167132.3167281"
}

@INPROCEEDINGS{Starke2022IntRS,
  title     = "Examining choice overload across single-list and multi-list user
               interfaces",
  author    = "Starke, Alain D and Sedkowska, Justyna and Chouhan, Mihir and
               Ferwerda, Bruce",
  booktitle = "Joint Workshop on Interfaces and Human
               Decision Making for Recommender Systems",
  pages     = "3--17",
  year      =  2022
}

@INPROCEEDINGS{Smets2022IntRS,
  title     = "Serendipity in recommender systems beyond the algorithm: A
               feature repository and experimental design",
  author    = "Smets, Annelien and Michiels, Lien and Bogers, Toine and
               Björneborn, Lennart",
  booktitle = "Proceedings of the 9th Joint Workshop on
               Interfaces and Human Decision Making for Recommender Systems
               co-located with 16th ACM Conference on Recommender Systems",
  pages     = "46--66",
  year      =  2022
}

@ARTICLE{Smets2023,
  title    = "Designing for serendipity: a means or an end?",
  author   = "Smets, Annelien",
  journal  = "Journal of Documentation",
  volume   =  79,
  number   =  3,
  pages    = "589--607",
  year     =  2023,
  doi      = "10.1108/JD-12-2021-0234"
}

@INPROCEEDINGS{Radle2012,
  title     = "{EBook} meets tabletop: Using collaborative visualization for
               search and serendipity in on-line book repositories",
  author    = "Rädle, Roman and Weiler, Andreas and Huber, Stephan and Jetter,
               Hans Christian and Mansmann, Svetlana and Reiterer, Harald and
               Scholl, Marc H",
  booktitle = "Proceedings of the fifth ACM workshop on
               Research advances in large digital book repositories and
               complementary media",
  pages     = "3--6",
  year      =  2012,
  doi       = "10.1145/2390116.2390120"
}

@BOOK{Race2016,
  title     = "Accidental information discovery: Cultivating serendipity in the
               digital age",
  author    = "Race, Tammera M and Makri, Stephann",
  publisher = "Chandos Publishing",
  year      =  2016
}

@INPROCEEDINGS{Parra2022,
  title     = "From user control and explainability in recommendation interfaces
               to Visual {XAI}",
  author    = "Parra, Denis",
  booktitle = "Joint Workshop on Interfaces and Human
               Decision Making for Recommender Systems",
  pages     = "1--2",
  year      =  2022
}

@INPROCEEDINGS{OkuK2011,
  title     = "Fusion-based recommender system for improving serendipity",
  author    = "Oku, Kenta and Hattori, Fumio",
  booktitle = "Workshop on Novelty and Diversity in
               Recommender Systems",
  pages     = "19--26",
  year      =  2011
}

@INPROCEEDINGS{OkuK2012,
  title     = "User evaluation of fusion-based approach for serendipity-oriented
               recommender system",
  author    = "Oku, Kenta and Hattori, Fumio",
  booktitle = "Workshop on Recommendation Utility
               Evaluation: Beyond RMSE",
  volume    =  910,
  pages     = "39--44",
  year      =  2012
}

@INPROCEEDINGS{MurakamiT2008,
  title     = "Metrics for evaluating the serendipity of recommendation lists",
  author    = "Murakami, Tomoko and Mori, Koichiro and Orihara, Ryohei",
  editor    = "Satoh, K and Inokuchi, A and Nagao, K and Kawamura, T",
  booktitle = "Proceedings of the 2007 conference on New frontiers in
               artificial intelligence",
  publisher = "Springer",
  volume    =  4914,
  pages     = "40--46",
  series    = "Lecture Notes in Computer Science",
  year      =  2008
}

@INPROCEEDINGS{Nguyen2014,
  title     = "Exploring the filter bubble: The effect of using recommender
               systems on content diversity",
  author    = "Nguyen, Tien T and Hui, Pik-Mai and Harper, F Maxwell and
               Terveen, Loren and Konstan, Joseph A",
  booktitle = "Proceedings of the 23rd international conference on
               World wide web",
  pages     = "677--686",
  year      =  2014,
  doi       = "10.1145/2566486.2568012"
}

@INPROCEEDINGS{Melo2018,
  title     = "The serendipitous pattern in interaction design",
  author    = "Melo, Ricardo and Carvalhais, Miguel",
  booktitle = "Proceedings of the 24th International Symposium on Electronic Art",
  pages     = "253--261",
  year      =  2018
}

@INPROCEEDINGS{McNee2006,
  title     = "Being accurate is not enough: How accuracy metrics have hurt
               recommender systems",
  author    = "McNee, Sean M and Riedl, John and Konstan, Joseph A",
  booktitle = "Extended Abstracts on Human Factors in Computing
               Systems",
  pages     = "1097--1101",
  year      =  2006
}

@ARTICLE{McCay-Peet2015IPM,
  title     = "Examination of relationships among serendipity, the environment,
               and individual differences",
  author    = "McCay-Peet, Lori and Toms, Elaine G and Kelloway, E Kevin",
  journal   = "Information Processing \& Management",
  publisher = "Elsevier Ltd",
  volume    =  51,
  number    =  4,
  pages     = "391--412",
  year      =  2015,
  doi       = "10.1016/j.ipm.2015.02.004"
}

@INPROCEEDINGS{Makri2015,
  title     = "Observing serendipity in digital information environments",
  author    = "Makri, Stephann and Bhuiya, Jaffor and Carthy, Jermaine and
               Owusu-Bonsu, Jermaine",
  booktitle = "Proceedings of the Association for Information Science and
               Technology",
  volume    =  52,
  pages     = "1--10",
  year      =  2015,
  doi       = "10.1002/pra2.2015.145052010019"
}

@ARTICLE{Makri2014,
  title    = "``making my own luck'': Serendipity strategies and how to support
              them in digital information environments",
  author   = "Makri, Stephann and Blandford, Ann and Woods, Mel and Sharples,
              Sarah and Maxwell, Deborah",
  journal  = "Journal of the Association for Information Science and Technology",
  volume   =  65,
  number   =  11,
  pages    = "2179--2194",
  year     =  2014,
  doi      = "10.1002/asi.23200"
}

@ARTICLE{LiuY2022,
  title    = "Serendipity in human information behavior: a systematic review",
  author   = "Liu, Yaxi and Qin, Chunxiu and Ma, Xubu and Liang, Huigang",
  journal  = "Journal of Documentation",
  volume   =  78,
  number   =  2,
  pages    = "435--462",
  year     =  2022,
  doi      = "10.1108/JD-02-2021-0029"
}

@INPROCEEDINGS{Kunkel2017,
  title     = "A {3D} item space visualization for presenting and manipulating
               user preferences in collaborative filtering",
  author    = "Kunkel, Johannes and Loepp, Benedikt and Ziegler, Jürgen",
  booktitle = "Proceedings of the 22nd International Conference on
               Intelligent User Interfaces",
  pages     = "3--15",
  year      =  2017,
  doi       = "10.1145/3025171.3025189"
}

@ARTICLE{Kotkov2016,
  title    = "A survey of serendipity in recommender systems",
  author   = "Kotkov, Denis and Wang, Shuaiqiang and Veijalainen, Jari",
  journal  = "Knowledge-Based Systems",
  volume   =  111,
  pages    = "180--192",
  year     =  2016,
  doi      = "10.1016/j.knosys.2016.08.014"
}

@INPROCEEDINGS{Kotkov2024CHIIR,
  title     = "The dark matter of serendipity in recommender systems",
  author    = "Kotkov, Denis and Medlar, Alan and Kask, Triin and Glowacka,
               Dorota",
  booktitle = "Proceedings of the 2024 Conference on Human
               Information Interaction and Retrieval",
  pages     = "108--118",
  year      =  2024,
  doi       = "10.1145/3627508.3638342"
}

@INPROCEEDINGS{Kotkov2023,
  title     = "Rethinking serendipity in recommender systems",
  author    = "Kotkov, Denis and Medlar, Alan and Glowacka, Dorota",
  booktitle = "Proceedings of the 2023 Conference on Human
               Information Interaction and Retrieval",
  pages     = "383--387",
  year      =  2023,
  doi       = "10.1145/3576840.3578310"
}

@INPROCEEDINGS{Khalili2017,
  title     = "Fostering serendipitous knowledge discovery using an adaptive
               multigraph-based faceted browser",
  author    = "Khalili, Ali and Van Andel, Pek and Van Den Besselaar, Peter and
               De Graaf, Klaas Andries",
  booktitle = "Proceedings of the 9th Knowledge Capture
               Conference",
  pages     = "1--4",
  year      =  2017,
  doi       = "10.1145/3148011.3148037"
}

@INPROCEEDINGS{Karger2004,
  title     = "Collections: Flexible, essential tools for information management",
  author    = "Karger, David R and Quan, Dennis",
  booktitle = "CHI '04 Extended Abstracts on Human Factors in
               Computing Systems",
  pages     = "1159--1162",
  year      =  2004,
  doi       = "10.1145/985921.986013"
}

@ARTICLE{Kaminskas2016,
  title    = "Diversity, serendipity, novelty, and coverage: A survey and
              empirical analysis of beyond-accuracy objectives in recommender
              systems",
  author   = "Kaminskas, Marius and Bridge, Derek",
  journal  = "ACM Transactions on Interactive Intelligent Systems",
  volume   =  7,
  number   =  1,
  pages    = "1--42",
  year     =  2016,
  doi      = "10.1145/2926720"
}

@INPROCEEDINGS{JinY2018RecSys,
  title     = "Effects of personal characteristics on music recommender systems
               with different levels of controllability",
  author    = "Jin, Yucheng and Tintarev, Nava and Verbert, Katrien",
  booktitle = "Proceedings of the 12th ACM Conference on Recommender
               Systems",
  pages     = "13--21",
  year      =  2018
}

@INPROCEEDINGS{Iaquinta2008,
  title     = "Introducing serendipity in a content-based recommender system",
  author    = "Iaquinta, Leo and de Gemmis, Marco and Lops, Pasquale and
               Semeraro, Giovanni and Filannino, Michele and Molino, Piero",
  booktitle = "2008 Eighth International Conference on Hybrid
               Intelligent Systems",
  publisher = "IEEE",
  pages     = "168--173",
  month     =  sep,
  year      =  2008,
  doi       = "10.1109/HIS.2008.25"
}

@ARTICLE{HeC2016,
  title     = "Interactive recommender systems: A survey of the state of the art
               and future research challenges and opportunities",
  author    = "He, Chen and Parra, Denis and Verbert, Katrien",
  journal   = "Expert Systems with Applications",
  publisher = "Elsevier Ltd",
  volume    =  56,
  pages     = "9--27",
  year      =  2016,
  doi       = "10.1016/j.eswa.2016.02.013"
}

@ARTICLE{Herlocker2004,
  title     = "Evaluating collaborative filtering recommender systems",
  author    = "Herlocker, Jonathan L and Konstan, Joseph A and Terveen, Loren G
               and Riedl, John T",
  journal   = "ACM Transactions on Information Systems",
  publisher = "ACM",
  volume    =  22,
  number    =  1,
  pages     = "5--53",
  year      =  2004
}

@INPROCEEDINGS{Harper2015RecSys,
  title     = "Putting users in control of their recommendations",
  author    = "Harper, F Maxwell and Xu, Funing and Kaur, Harmanpreet and
               Condiff, Kyle and Chang, Shuo and Terveen, Loren",
  booktitle = "Proceedings of the 9th ACM Conference on
               Recommender Systems",
  pages     = "3--10",
  year      =  2015
}

@INPROCEEDINGS{Hangal2012,
  title     = "Effective browsing and serendipitous discovery with an
               experience-infused browser",
  author    = "Hangal, Sudheendra and Nagpal, Abhinay and Lam, Monica S",
  booktitle = "Proceedings of the 2012 ACM international conference
               on Intelligent User Interfaces",
  pages     = "149--158",
  year      =  2012,
  doi       = "10.1145/2166966.2166994"
}

@INPROCEEDINGS{Green2009,
  title     = "Generating transparent, steerable recommendations from textual
               descriptions of items",
  author    = "Green, Stephen J and Lamere, Paul and Alexander, Jeffrey and
               Maillet, François and Kirk, Susanna and Holt, Jessica and
               Bourque, Jackie and Mak, Xiao-Wen",
  booktitle = "Proceedings of the third ACM conference on
               Recommender systems",
  pages     = "281--284",
  year      =  2009,
  doi       = "10.1145/1639714.1639768"
}

@INPROCEEDINGS{Gansner2009,
  title     = "Putting recommendations on the map: Visualizing clusters and
               relations",
  author    = "Gansner, Emden and Hu, Yifan and Kobourov, Stephen and Volinsky,
               Chris",
  booktitle = "Proceedings of the third ACM conference on
               Recommender systems",
  pages     = "345--348",
  year      =  2009,
  doi       = "10.1145/1639714.1639784"
}

@ARTICLE{Erdelez2016,
  title    = "Research perspectives on serendipity and information encountering",
  author   = "Erdelez, Sanda and Heinström, Jannica and Makri, Stephann and
              Björneborn, Lennart and Beheshti, Jamshid and Toms, Elaine and
              Agarwal, Naresh Kumar",
  journal  = "Proceedings of the Association for Information Science and
              Technology",
  volume   =  53,
  number   =  1,
  pages    = "1--5",
  year     =  2016,
  doi      = "10.1002/pra2.2016.14505301011"
}

@INPROCEEDINGS{Dias2014HCI,
  title     = "Interactive visualization for music rediscovery and serendipity",
  author    = "Dias, Ricardo and Pinto, Joana and Fonseca, Manuel J",
  booktitle = "Proceedings of the 28th International BCS Human
               Computer Interaction Conference",
  pages     = "183--188",
  year      =  2014,
  doi       = "10.14236/ewic/hci2014.20"
}

@INPROCEEDINGS{ChenL2019WWW,
  title     = "How serendipity improves user satisfaction with recommendations?
               A large-scale user evaluation",
  author    = "Chen, Li and Yang, Yonghua and Wang, Ningxia and Yang, Keping and
               Yuan, Quan",
  booktitle = "The World Wide Web Conference",
  pages     = "240--250",
  year      =  2019,
  doi       = "10.1145/3308558.3313469"
}

@INCOLLECTION{Castells2015,
  title     = "Novelty and diversity in recommender systems",
  author    = "Castells, Pablo and Hurley, Neil J and Vargas, Saul",
  editor    = "Ricci, Francesco and Rokach, Lior and Shapira, Bracha",
  booktitle = "Recommender Systems Handbook",
  publisher = "Springer",
  edition   =  2,
  pages     = "881--918",
  year      =  2015,
  doi       = "10.1007/978-1-4899-7637-6\_26"
}

@INPROCEEDINGS{Burkell2012,
  title     = "Promoting serendipity online: Recommendations for tool design",
  author    = "Burkell, Jacquelyn and Quan-Haase, Anabel and Rubin, Victoria L",
  booktitle = "Proceedings of the 2012 iConference",
  pages     = "525--526",
  year      =  2012,
  doi       = "10.1145/2132176.2132274"
}

@INPROCEEDINGS{Bollen2010,
  title     = "Understanding choice overload in recommender systems",
  author    = "Bollen, Dirk and Knijnenburg, Bart P and Willemsen, Martijn C and
               Graus, Mark",
  booktitle = "Proceedings of the fourth ACM conference on
               Recommender systems",
  pages     = "63--70",
  year      =  2010
}

@ARTICLE{Bjorneborn2017,
  title    = "Three key affordances for serendipity: Toward a framework
              connecting environmental and personal factors in serendipitous
              encounters",
  author   = "Björneborn, Lennart",
  journal  = "Journal of Documentation",
  volume   =  73,
  number   =  5,
  pages    = "1053--1081",
  year     =  2017,
  doi      = "10.1108/JD-07-2016-0097"
}

@ARTICLE{Van-Andel1994,
  title    = "Anatomy of the unsought finding. Serendipity: Origin, history,
              domains, traditions, appearances, patterns and programmability",
  author   = "Van Andel, Pek",
  journal  = "British Journal for the Philosophy of Science",
  volume   =  45,
  number   =  2,
  pages    = "631--648",
  year     =  1994,
  doi      = "10.1093/bjps/45.2.631"
}

@ARTICLE{Afridi2023,
  title     = "Review of user interface-facilitated serendipity in recommender
               systems",
  author    = "Afridi, Ahmad Hassan and Olsson, Thomas",
  journal   = "International Journal of Interactive Communication
               Systems and Technologies",
  publisher = "IGI Global",
  volume    =  12,
  number    =  1,
  pages     = "1--19",
  month     =  mar,
  year      =  2023,
  doi       = "10.4018/ijicst.320180"
}

@ARTICLE{Afridi2018b,
  title     = "Visualizing serendipitous recommendations in user controlled
               recommender system for learning",
  author    = "Afridi, Ahmad Hassan",
  journal   = "Procedia Computer Science",
  publisher = "Elsevier B.V.",
  volume    =  141,
  pages     = "496--502",
  year      =  2018,
  doi       = "10.1016/j.procs.2018.10.136"
}

@ARTICLE{Afridi2018a,
  title     = "User control and serendipitous recommendations in learning
               environments",
  author    = "Afridi, Ahmad Hassan",
  journal   = "Procedia Computer Science",
  publisher = "Elsevier B.V.",
  volume    =  130,
  pages     = "214--221",
  year      =  2018,
  doi       = "10.1016/j.procs.2018.04.032"
}

@INPROCEEDINGS{Bostandjiev2012,
  title     = "{TasteWeights}: A visual interactive hybrid recommender system",
  author    = "Bostandjiev, Svetlin and O'Donovan, John and Höllerer, Tobias",
  booktitle = "Proceedings of the sixth ACM conference on
               Recommender systems",
  publisher = "ACM",
  address   = "New York, NY, USA",
  pages     = "35--42",
  month     =  "9~" # sep,
  year      =  2012,
  doi       = "10.1145/2365952.2365964",
  language  = "en"
}

@INPROCEEDINGS{Andjelkovic2016,
  title     = "Moodplay: Interactive mood-based music discovery and
               recommendation",
  author    = "Andjelkovic, Ivana and Parra, Denis and O'Donovan, John",
  booktitle = "Proceedings of the 2016 Conference on User Modeling
               Adaptation and Personalization",
  publisher = "ACM",
  address   = "New York, NY, USA",
  pages     = "275--279",
  month     =  "13~" # jul,
  year      =  2016,
  doi       = "10.1145/2930238.2930280",
  language  = "en"
}


\end{document}